%% file: lenses.tex
\input mn

\input mnextra

\input psfig

\overfullrule=0pt
\newif\ifpsfiles\psfilesfalse
\def\getfig#1#2{\ifpsfiles\psfig{figure=#1,width=6.5cm}\else\vskip#2cm\fi}
\def\getfigh#1#2{\ifpsfiles\psfig{figure=#1,width=\hsize}\else\vskip#2cm\fi}
\hyphenation{dimen-sional}

\def\ex#1{\langle#1\rangle}
\def\micron{\,\mu\hbox{m}}
\def\etal{et al.}

\def\apj #1 #2 {ApJ, #1, #2}
\def\apjsupp #1 #2 {ApJSuppl, #1, #2}
\def\aj #1 #2 {AJ, #1, #2}
\def\mn #1 #2 {MNRAS, #1, #2}
\def\aa #1 #2 {A\&A, #1, #2}

\Autonumber\begintopmatter
\title{Microlensing Optical Depth of the COBE Bulge}

\author{Nicolai Bissantz$^{1,2}$, Peter Englmaier$^1$,
James Binney$^3$, Ortwin Gerhard$^1$}

\affiliation{$^1$Astronomisches Institut, Universit\"at Basel,
Venusstrasse 7, CH-4102 Binningen, Switzerland}
\vskip4pt
\affiliation{$^2$Physikalisches Institut, Universit\"at Freiburg,
Herrmann-Herder-Strasse 3, D-79104 Freiburg, Germany}
\vskip4pt
\affiliation{$^3$Theoretical Physics, Keble Road, Oxford OX1 3NP}
\vskip4pt

\shortauthor{N.~Bissantz, P.~Englmaier, J.J.~Binney, O.E.~Gerhard}
\shorttitle{Optical depth of COBE bulge}

\acceptedline{Accepted xx. Received xx}

\abstract{We examine the left--right asymmetry in the
cleaned COBE/DIRBE near-infrared data of the inner Galaxy and show (i)
that the Galactic bar is probably not seen very nearly end-on, and
(ii) that even if it is, it is not highly elongated. The assumption of
constant mass-to-light ratio is used to derive simulated
terminal--velocity plots for the ISM from our model luminosity
distributions. By comparing these plots with observed terminal
velocities we determine the mass-to-light ratio of the near-IR bulge
and disk.

Assuming that all this mass contributes to gravitational microlensing
we compute optical depths $\tau$ for microlensing in Galactic-centre
fields.  For three models with bar major axis between $10\deg-25\deg$
from the Sun-Galactic Center line, the resulting optical depths in
Baade's window lie in the range $0.83\times10^{-6}\lta\tau\lta
0.89\times10^{-6}$ for main-sequence stars and
$1.2\times10^{-6}\lta\tau\lta 1.3\times10^{-6}$ for red-clump
giants. We discuss a number of uncertainties including possible
variations of the near-infrared mass-to-light ratio. We conclude that,
although the values predicted from analyzing the COBE and gas velocity
data are inconsistent at the $2-2.5\sigma$ level with recent
observational determinations of $\tau$, we believe they should be
taken seriously.}


\maketitle

\section{Introduction}

It is widely recognized that observed optical depths to microlensing
along various lines of sight from the Sun constitute important
constraints on models of the Milky Way. In his seminal study of
microlensing Paczynski (1986) estimated optical depths $\tau$ for
lensing under the assumption that the Galaxy is axisymmetric. He found
that in fields towards the Galactic centre $\tau$ is $\sim10^{-6}$,
and numerous studies have since confirmed the correctness of this
estimate.  When $\tau$ was actually measured, significantly larger
values have been reported -- $\tau=(3.3\pm1.2)\times10^{-6}$
($1\sigma$) by (Udalski et al (1994) and
$\tau=(1.9\pm0.4)\times10^{-6}$ by Alcock \etal\ (1995)].  It was
suggested (Paczynski \etal\ 1994; Zhao, Spergel \& Rich 1995) that the
large measured values of $\tau$ reflect the fact that the Galaxy is
barred (see, e.g., reviews in Blitz \& Teuben 1996).

In principle a suitably oriented and elongated bar can significantly enhance
$\tau$. To see this, imagine deforming an initially axisymmetric bulge into
a bar that points nearly to the Sun in such a way as to hold constant the
velocity dispersion within Baade's window [$(l,b)=(1\deg,-3.9\deg)$]. This
deformation will increase both the surface density of stars towards Baade's
window and the depth along this line of sight within which the stellar
density is high. Since a star offers the largest cross section for lensing a
background star when it lies halfway between the background star and us, both
of these factors will increase $\tau$. However, in this paper we show
that if one assumes that near-infrared luminosity density is a fair tracer
of stellar mass density in the inner galaxy, then $\tau$ is rather precisely
constrained to be $\tau=10^{-6}$ by a combination of (i) the
near and far-infrared surface brightnesses that were measured by the DIRBE
experiment aboard the COBE satellite, and (ii) the kinematics of gas in the
inner few kiloparsecs as reflected in longitude--velocity plots from
radio-frequency emission-line studies of the ISM.

In substantiating our claim we rely heavily on results obtained in three
previous papers, namely Binney \& Gerhard (1996; Paper I), Spergel, Malhotra
\& Blitz (1997; Paper II) and Binney, Gerhard \& Spergel (1997; Paper III).
Paper II cleans the COBE/DIRBE data for the effects of obscuration by dust,
while Papers I and III develop and apply a Lucy--Richardson algorithm for
recovering models of the Galaxy's three-dimensional luminosity density from
the cleaned near-IR surface photometry.

This paper is organized as follows. In Section 2 we focus on
left--right asymmetry within the observed near-IR surface brightness
measurements as a key diagnostic of the morphology of the Galactic
bar. In particular we explain how the length, axis ratios and
orientation of a bar are reflected in its distribution of left--right
asymmetry. We then compare these asymmetries both with the asymmetry
that is apparent in the COBE data, and the asymmetries that are
predicted by our models of the Galaxy. These comparisons suggest that
the Galactic bar is not seen very nearly end-on, and even if it is, it
is not highly elongated.

In Section 3 we assume that our models of the near-IR luminosity
density can be converted into mass models by multiplying by an
appropriate mass-to-light ratio $\Upsilon$. We calibrate $\Upsilon$ by
comparing simulated longitude--velocity plots of the ISM with observed
$(l,v)$ plots. Finally we determine $\tau$. In Section 4 we discuss
uncertainties and the possible causes for the difference between the
optical depth inferred here and that inferred from the microlensing
experiments.

\section{Left--right asymmetry}

As Blitz \& Spergel (1991) have emphasized, the key to using photometry to
detect departures from axisymmetry within the Galaxy is the study of the
ratio $\Delta(l,b)$ that is defined for positive $l$ by
 $$
\Delta(l,b)\equiv I(l,b)/I(-l,b),
\eqno\newe$$
 where $I(l,b)$ is the surface brightness of the Galaxy at longitude $l$ and
latitude $b$. Fig.~1 is a contour plot of $\Delta$ for the $L$-band COBE
data after the latter has been corrected for dust obscuration as described
in Spergel, Malhotra \& Blitz (1996).  Values of $\Delta$ are available for
pixels that are $1.5\deg$ square, but the values contoured in Fig.~1 are
those obtained by smoothing the raw data by 5 pixels in $l$ and 2 pixels in
$b$ with the {\tt SMOOFT} routine of Press \etal\ (1986). We shall call a plot
such as Fig.~1 an {\it asymmetry map}.

\beginfigure1
\getfig{diffobs.ps}{6.5}
\caption{{\bf Figure 1.} The asymmetry map of the COBE $L$-band data.
Contours are spaced by $0.05\,$magnitudes. Dotted contours indicate that the
Galaxy is brighter on the right than on the left.}
\endfigure

\begintable1
\caption{{\bf Table 1.} Parameters of the models}
\tabskip=1em plus 2em minus 0.5em%
\halign to 6cm{
\hfil$#$\hfil&\hfil$#$\hfil&\hfil$#$\hfil&\hfil$#$\hfil&\hfil$#$\hfil&\hfil$#$\hfil\cr
\noalign{\vskip2pt\hrule\vskip2pt\hrule\vskip2pt}
a_0&R_\d&\alpha&z_0&z_1&B_0/a_{\rm m}^3\eta\zeta\cr
\noalign{\vskip2pt\hrule\vskip2pt}
100\pc&2.5\kpc&0.27&210\pc&42\pc&624\hfil\cr
\noalign{\vskip2pt\hrule\vskip2pt}
}
\endtable

\beginfigure*2
\ifpsfiles
\centerline{\psfig{figure=da25thin.ps,width=5cm}\hfil
\psfig{figure=da25dens.ps,width=5cm}}
\centerline{\psfig{figure=da25.ps,width=5cm}\hfil
\psfig{figure=da25long.ps,width=5cm}}
\centerline{\psfig{figure=da10.ps,width=5cm}\hfil
\psfig{figure=da10thin.ps,width=5cm}}
\fi
\caption{{\bf Figure 2.} Asymmetry maps of analytical models. The
parameters in eqs (2) of each model are given at the top of its panel.
The contours are as in Fig.~1.}
\endfigure

\beginfigure*3
\ifpsfiles
\centerline{\psfig{figure=d25thin.ps,width=6cm}\hfil
\psfig{figure=d25dens.ps,width=6cm}}
\centerline{\psfig{figure=d25.ps,width=6cm}\hfil
\psfig{figure=d25long.ps,width=6cm}}
\centerline{\psfig{figure=d10.ps,width=6cm}\hfil
\psfig{figure=d10thin.ps,width=6cm}}
\fi
\caption{{\bf Figure 3.} Asymmetry maps of iterated models. The parameters
in eqs (2) of the model from which the iterations started are given at the
top of each panel. The contours are as in Fig.~1.}
\endfigure

To appreciate the significance of Fig.~1 it is instructive to construct
analogous figures for simple Galaxy models. Fig.~2 shows several illustrative
cases for the  model introduced in Paper I. This consists of a bar/bulge and
a disk. Its luminosity density is given by
 \eqnam\analfit$$
j(\b x)=j_0 \left[ f_{\rm b}(\b x)+f_\d(\b x) \right],
\eqno\firste$$
 where
$$\eqalign{
f_{\rm b}&\equiv{B_0\over a_{\rm m}^3\eta\zeta}{\e^{-a^2/a_{\rm m}^2}\over(1+a/a_0)^{1.8}}\cr
f_\d&\equiv\Big({\e^{-|z|/z_0}\over z_0}+\alpha{\e^{-|z|/z_1}\over
z_1}\Big)R_\d \e^{-R/R_\d}\cr 
a&\equiv\bigg(x^2+{y^2\over \eta^2}+{z^2\over
\zeta^2}\bigg)^{1/2}\quad\hbox{and}\quad
R\equiv(x^2+y^2)^{1/2}.  }
\eqno\laste b$$ 
  There are four important parameters associated with this model bar:
the length $a_{\rm m}$, the axis ratios $\eta,\,\zeta$ and the angle
$\phi_0$ between the Sun--centre line and the long axis of the bar
(which is defined such that the near end of the bar lies at $l>0$ for
$0<\phi_0<90\deg$). The other parameters may be held constant at the
values given in Table 1. We assume throughout that the Sun lies
$14\pc$ below the Galactic plane as was deduced in Paper III.

\beginfigure*4
\ifpsfiles
\centerline{\psfig{figure=pr25.ps,width=5.8cm}\hfil
\psfig{figure=pr25long.ps,width=5.8cm}\hfil
\psfig{figure=pr10thin.ps,width=5.8cm}}
\fi
\caption{{\bf Figure 4.} Projections parallel to $z$ of three final models:
From left to right: for $\phi_0=25\deg$ started from a short-fat analytic
bar (middle left panel of Fig.~2), and started from a longer bar of similar
cross-section (middle-right panel of Fig.~2); for $\phi_0=10$ started from a
long thin bar (lower right panel of Fig.~2).}
\endfigure

The top-left panel of Fig.~2 shows the asymmetry map for $a_{\rm m}=1.91\kpc$,
$\eta=0.5$, $\zeta=0.3$, $\phi_0=25\deg$. In contrast to the data map, Fig.~1,
all contours are positive, that is, for the given parameters, the model
\refe1) is nowhere more than $0.05\,$mag brighter at $l>0$ than it is at
$l<0$. The brightness contrast peaks near $(l,b)=(8\deg,\pm4\deg)$. For the
middle-left panel of Fig.~2 the vertical axis ratio $\zeta$ has been doubled
to $\zeta=0.6$. This moves the peaks in the asymmetry map out away from $b=0$
and increases their intensity. The top-right panel in Fig.~2 shows the
effect of decreasing the planar axis ratio from $0.5$ to $\eta=0.3$ whilst
holding the vertical axis ratio constant at $\zeta=0.3$.  This leaves the
peaks at approximately the same location but makes them more intense. The
middle-right panel of Fig.~2 shows the effect of doubling the length of the
bar from $1.91\kpc$ to $3.82\kpc$ whilst holding the axis ratios constant at
$0.3$.  This both moves the peaks towards larger $l$ and higher $b$ and
makes the surrounding contours significantly more extensive. Comparison of
the middle and bottom-left panels of Fig.~2 shows the effect rotating the bar
to a more nearly end-on orientation; the bottom-left panel is for
$\phi_0=10\deg$ rather than $\phi_0=25\deg$. This leaves the peaks in
approximately the same location because even at $\phi_0=25\deg$ the line of
sight through the Galactic centre effectively passes down the length of this
fairly fat bar rather than intersecting it transversely. In this regime the
magnitude of the peaks decreases with $\phi_0$ -- by symmetry it must vanish
at $\phi_0=0$. 

The bottom-right panel of Fig.~2 shows the asymmetry map of a long
($a_{\rm m}=3.82\kpc$) thin ($\eta=0.25,\zeta=0.2$) bar seen nearly end-on
($\phi_0=10\deg$). Comparison of this asymmetry map with that shown in the
middle-right panel of Fig.~2 reveals that the effect of rotating a long thin
bar to a more nearly end-on orientation is two-fold: it decreases the
magnitude of the peaks as in the case of a short bar, but it now also shifts
them to slightly lower longitudes.  This is because when a thin bar is
oriented at $\phi_0=25\deg$, the line of sight through the Galactic centre
intersects the bar transversely.  Finally, comparison of the two lower-right
panels with the corresponding lower-left panels shows that rotation of a bar
towards end-on orientation combined with elongation of the bar can leave the
peaks in the asymmetry map at approximately the same locations.  However
this combination of rotation and elongation modifies the shapes of the
contours that surround the peaks in a characteristic way: they become more
elongated in the longitude direction, and they turn upward towards high
latitudes at $l\gta 20\deg$. This effect arises because the far end of a
long near end-on bar appears small because it is far off, so away from the
plane and at large longitudes there is no counterpoint to the brightness
contributed by the apparently large near end of the bar. Conversely, at
small $l$ and $b$ light from the far end of the bar provides an effective
counterpoint to light from the near end of the bar, and the asymmetry is
small.

Fig.~3 shows the asymmetry maps of the models that are generated by the
algorithm of Papers I and III from the COBE $L$-band data when the
Lucy--Richardson iterations are started from the analytic models that
generate the asymmetry maps of the corresponding panels in Fig.~2. Several
points are noteworthy  in this figure:

(i) Especially in the $\phi_0=25\deg$ case, the asymmetry map of the model
generated by the algorithm is remarkably independent of the initial analytic
model, even though the length and axis ratios of the latter vary by factors
of two.

(ii) The $\phi_0=25\deg$ final model provides a remarkably good fit to
the global morphology of the COBE asymmetry map of Fig.~1. In
particular, in the middle-right panel of Fig.~3 the two long ridges of
maximal asymmetry around $(l,b)=(10\deg,\pm6\deg)$ strongly resemble
ridges in Fig.~1 that have similar locations and orientations.
Moreover, the height of these ridges is nearly the same in the two
figures, as is the general shape of the low-lying contours near
$l=20\deg$. Finer details of the observed asymmetry map shown in
Fig.~1 are not reproduced in Fig.~3. These include the sharp local
maximum at $(l,b)=(16\deg,-1\deg)$ and the strong north-south
asymmetry.  The former is almost certainly associated with a feature
in the disk and does not concern us here.  The north--south asymmetry
in the observed asymmetry map could be a sign that the bar is slightly
inclined with respect to the plane, as radio-frequency emission lines
from the ISM suggest (e.g., Liszt \& Burton 1996).

(iii) The asymmetry maps that are shown in the bottom panels of Fig.~3 for
the case $\phi_0=10\deg$ both deviate from the observed asymmetry map of
Fig.~1 in having contours at $(l,b)\simeq(25\deg,\pm7\deg)$ which slope
towards higher $l$ and $|b|$ rather than towards lower $l$ at higher $|b|$.
Thus whereas in the case of $\phi_0=25\deg$ the Lucy--Richardson iterations
are able to eliminate the extensions towards high $l$ and $|b|$ that are
apparent in the middle-right panel of Fig.~2, the iterations cannot
eliminate this unsatisfactory feature of the initial models in the case
$\phi_0=10\deg$. When a bar is viewed nearly end-on it is inevitable that the
sky is brighter up and to the left than it is at the corresponding point
on the right.

\section{Three-dimensional structures and mass densities}

Fig.~4 shows projections parallel to $z$ of the bars that were
generated by six Lucy--Richardson iterations on the COBE $L$-band data
from the analytic bars whose asymmetry maps are shown in the
middle-left, middle-right and bottom-right panels of Fig.~2. 
%
These bars are very similar to one another. In particular, the
$25\deg$-model that started with a long bar finishes (middle panel)
with a bar of exactly the same length as the $25\deg$-model that
started with a bar of half the length (left-hand panel).  Comparing
the middle two panels of Figs 2 and 3, it is clear that the
Lucy--Richardson iterations have dealt with the problem posed by the
extension towards large $l$ and $|b|$ of the peaks in the middle-right
panel of Fig.~2 by shortening the bar. The $10\deg$-model shown in the
right-hand panel of Fig.~4 has a slightly {\it shorter\/} bar than do
the $25\deg$-models shown in the middle and left-hand panels. Thus
again the Lucy--Richardson iterations have shortened the bar, but the
bottom-right panel of Fig.~3 shows that now shortening is not
accompanied by elimination of the unwanted extensions of the asymmetry
peaks towards high $l$ and $|b|$. On the other hand, shortening the
bar has made it possible to broaden it, and altogether this has
increased the amplitudes of the peaks (cf.\ Fig.~2), as the
observations demand.  These experiments demonstrate that the gross
structure of the models recovered by Lucy--Richardson iteration can be
understood in simple general terms. Hence we may confidently assert
that the Galactic surface-brightness distribution is such that the
length of the bar that is required to fit the COBE data for given
$\phi_0$ is nearly independent of $\phi_0$ for $\phi\lta30\deg$. We
show below that the same is true of the optical depth to microlensing.

Before an optical depth can be determined, it is necessary to
associate a mass-to-light ratio $\Upsilon$ with a model. This we do as
follows. We first construct a luminosity model for the entire Galaxy
out to the solar radius, by combining the luminosity distribution
recovered by the Richardson--Lucy algorithm inside our $5\kpc \times
5\kpc \times 1.4\kpc$ computational box with the initial model of
Paper III in the region outside the box.  Then we reconstruct the cusp
in the centre, which is unresolved in the DIRBE data, by fitting a
power law to the multipole expansion of the luminosity distribution in
the range $350 - 500 \pc$.  We next evaluate the combined model's
potential under the assumption of constant $\Upsilon$. Then we use an
SPH code to simulate the flow of gas inside $\sim8\kpc$ [see Englmaier
\& Gerhard (1997) for details], including a higher resolution
simulation for the central $\sim4\kpc$. In this simulation, the
assumed pattern speed of the bar is $60 \kms /\kpc$, placing
corotation at $3.1\kpc$.  The gas flow velocities in the simulation
are transformed into a putative local standard of rest (LSR) frame at
the position of the Sun at $R_0=8\kpc$. For this transformation we
assume that the LSR has a tangential velocity $v_0$, but no component
of motion in the direction of the Galactic center. For each
line-of-sight we evaluate the maximum velocity that would be seen in
radio-frequency emission-line measurements.

We compare the resulting model terminal-velocity curve with the
observed HI and CO terminal velocities (corrected for line width, but
not for peculiar LSR motion; Burton \& Liszt 1993, Clemens
1985). Specifically, for $v_0=180,190,200,210,220\kms$ we choose the
value of $\Upsilon$ that gives the best eye-ball fit to the observed
velocities in the region between $l=17-48\deg$. Finally we choose an
optimal $(v_0,\Upsilon)$ combination.

\beginfigure5
\getfigh{termc.ps}{6}
 \caption{{\bf Figure 5.} Comparison of observed and model terminal
velocities. The diamonds show the terminal velocities and their error
bars from the CO data of Clemens (1985). The squares show unpublished
HI terminal velocity measurements of Burton. The stars are HI terminal
velocities read from the figures in Burton \& Liszt (1993).  The
curves show the predictions from three SPH gas flow models after
subtracting the component of solar tangential motion, $v_0 \sin l$,
with $v_0 = 200\kms$. Thin full curve: gas flow in the mass
distribution obtained by inverting the COBE surface brightness map
with $\phi_0=20\deg$, at intermediate numerical resolution.  Thick
full curve: the same model in a higher resolution simulation.  Dotted
curve: gas flow in model inverted with $\phi_0=10\deg$. The Sun's
galactocentric radius has been taken to be $R_0=8\kpc$. The models
have been observed at time $t=0.12\Gyr$ and have been scaled to fit by
eye the observed data between $l=17-48\deg$.}
\endfigure

The thin full curve in Fig.~5 shows the fit between model and data for
a model with $\phi_0=20\deg$ under the assumption $v_0=200\kms$ -- the
fit to the data at $(17\deg<l<48\deg)$ is consistent with the scatter
in the data points ($\sim 5\%$).  In the range $(9\deg\lta l \lta
17\deg)$ the model curve falls below the data by $\lta20\kms$. This
may reflect that fact that the photometric inversions show a strongly
elliptic disk at $(1.5\kpc\lta r\lta3.5\kpc)$, and, as discussed in
Paper III, this could be caused by the presence of supergiants in
star-forming regions along prominent spiral arms.  If so, our
assumption of constant $\Upsilon$ may not be correct in
this region and the gravitational forces in our models may be
insufficiently accurate at radii corresponding to $(9\deg\lta l \lta
17\deg)$.  At $l\sim3\deg$ the thin, full model curve in Fig.~5 falls
below the sharply peaked data. This may in part be due to the 
model's pattern speed being not quite correct, but in large
measure it undoubtedly reflects the limited spatial resolution of our
models: the thick, full curve in Fig.~5 shows the predictions for
exactly the same model as the thin, full curve, but calculated from a
higher-resolution SPH simulation. It can be seen that the thick curve
halves the shortfall of the thin curve with respect to the data.
However, this gas flow is not stationary, and as more gas falls
towards the Galactic center and the vicinity of the cusped orbit
is depopulated, the maximum velocities seen in the bulge region
decrease again.

The dotted curve in Fig.~5 shows the predictions of the model with
$\phi_0=10\deg$. Again $v_0=200\kms$ has been assumed. The match to
the data for this model is reasonable but somewhat inferior to that
obtained with the corresponding $\phi_0=20\deg$ model (thin, full
curve).  Specifically, for $\phi_0=10\deg$ there are systematic
deviations $\sim 10\kms$ from the observed velocities. About half the
difference between the terminal velocity curves of the $10\deg$ and
$20\deg$ models is due to the difference in viewing angles; the
remaining discrepancy is due to intrinsic differences in the model gas
flows.

The quality of the fit between the thick, full curve and the data in Fig.~5
indicates that to first order we may assume that the mass-to-light ratios of
the bulge and disk are equal. With this assumption we find
 \eqnam\rhob
$$ \rho(\b x) =\Upsilon_L\, j_L(\b x),
\eqno\firste$$
 where $j_L(\b x)$ is in COBE/DIRBE luminosity units per $\kpc^3$ as in
Paper III and 
$$\eqalign{
\Upsilon_L(\phi_0=20\deg)&=3.47\times10^8\msun/\hbox{COBE unit},\cr
\Upsilon_L(\phi_0=10\deg)&=3.37\times10^8\msun/\hbox{COBE unit}.
}\eqno\laste b$$
 Thus the derived mass-to-light ratio $\Upsilon_L$ is almost
independent of $\phi_0$. 

The uncertainties in the normalization of $\rho(\b x)$ are as
follows: Changing the tangential velocity of the LSR to $180\kms$ and
$220\kms$ changes $\Upsilon_L$ by $-10\%$ and $+10\%$, respectively.
From the above discussion of the velocity peak in Fig.~5, we conclude
that the near-IR mass-to-light ratio of the bulge can differ from that
of the disk by at most $\lta 20\%$.  Other uncertainties such as that
caused by the unknown projection angle $\phi_0$ are smaller.

In Table 2 we give approximate bulge masses interior to cylindrical radius
$R=2.4\kpc$ for the three COBE models of Fig.~4. These were estimated from
the complete density field $\rho(\b x)$ in the following way.  We associated
with the disk all mass that lies within $0.1$ radian of the plane (as seen
from the Galactic centre).  For each bounding radius $R$ we compared the
total mass $M_{\rm tot}(R)$ inside $R$ with the mass $M_{\rm b}(R)$ inside
$R$ that is, by the above definition, non-disk mass. We found that $M_{\rm
b}$ saturates for $R$ between $2-3\kpc$. For $R\lta2.5\kpc$ the difference
$M_{\rm tot}-M_{\rm b}$ is approximately equal to the mass of the initial
analytic disk model of Paper III that lies within $R$, while further out
there is significant disk mass that is not accounted for by the initial
analytic model. From these results we infer that a reasonable estimate of
the bulge's mass is given by the difference between $M_{\rm tot}(2.4\kpc)
= 1.9\times 10^{10} \msun$ and the mass of the initial disk that lies at
$R\le2.4\kpc$. This is the quantity that is given for each model in Table 2.

\section{Microlensing optical depths}

\begintable2
\caption{{\bf Table 2.} Microlensing optical depths in Baade's window
for the three COBE models shown in Fig.~6, in units of $10^{-6}$.}
\tabskip=1em plus 2em minus 0.5em%
\halign to 6.5cm{
\hfil#\hfil&\hfil$#$\hfil&\hfil$#$\hfil&\hfil$#$\hfil
&\hfil$#$\hfil&\hfil$#$\hfil\cr
\noalign{\vskip2pt\hrule\vskip2pt\hrule\vskip2pt}
Model&\phi_0&a_{\rm m}&M_{b,2.4}&\tau_{-6}^{\beta=0}
&\tau_{-6}^{\beta=-1}\cr
\noalign{\vskip2pt\hrule\vskip2pt}
1&25\deg&1.91\kpc&8.6\times 10^9&1.2&0.86\cr	
2&25\deg&3.82\kpc&8.1\times 10^9&1.2&0.83\cr	
3&10\deg&3.82\kpc&7.2\times 10^9&1.3&0.86\cr	
\noalign{\vskip2pt\hrule\vskip2pt}
}
\endtable

\beginfigure6
\getfigh{tauminor.ps}{6}
 \caption{{\bf Figure 6.} Microlensing optical depth for several
models along the minor axis ($l=0$). The short-dashed lines show
optical depth as a function of latitude for the
$\phi_0=25\deg$-bulge-disk model (top-left panel of Fig.~3). The
long-dashed lines show optical depths for the long
$\phi_0=25\deg$-model (middle-right panel of Fig.~3 and middle panel
of Fig.~4). The solid lines show optical depths for the
$\phi_0=10\deg$-model (bottom-right panel of Fig.~3 and right panel of
Fig.~4). For all models, the upper curve is for $\beta=0$ and the
lower curve is for $\beta=-1$.}
\endfigure

\beginfigure7
\getfigh{taumap.ps}{6}
 \caption{{\bf Figure 7.} Microlensing optical depth map for the model
shown in the middle-right panel of Fig.~3 and middle panel of
Fig.~4. The thick contour has $\tau_{-6}=1$; the spacing between
contours corresponds to a factor of $1.5$.}
\endfigure

The microlensing optical depth $\tau(D_s)$ for a source at distance $D_s$
is the probability for it to fall within one Einstein radius of any
intervening star:
$$ \eqalign{
	\tau(D_{\rm s})=&\int_0^{D_{\rm s}}\!{4\pi G\rho(D_\d) \over c^2}
			{D_\d D_{ds} \over D_{\rm s}}\,\d D_\d \cr
	 =& {4\pi G\over c^2} D_{\rm s}^2 \int_0^1 \rho(x) x(1-x)\,\d x,\cr}
\eqno\newe$$
 where $\rho(D_\d)$ is the mass density of lenses at distance $D_\d$,
$x=D_\d/D_{\rm s}$, and $D_{\rm s}=D_\d+D_{\rm ds}$. $\tau$ depends only on
the mass density of lenses, and is independent of the lensing mass spectrum.
The measured optical depth in the direction $(l,b)$ is the function
$\tau(D_{\rm s})$ averaged over all observable sources that are brighter
than some apparent magnitude $m$ in a cone of small $\delta l,\delta b$
around ($l,b$).  Following Kiraga \& Paczynski (1994) it is customary to
parameterize the distribution in $D_{\rm s}$ of these sources as 
 $$
{\d N(D_{\rm s})\over\d D_{\rm s}}=\hbox{constant}\times
\rho(D_{\rm s})D_{\rm s}^{2+2\beta}.
\eqno\newe$$
 This assumes that the luminosity function of sources is constant along the
line-of-sight and that the fraction of stars brighter than luminosity $L$ is
proportional to $L^{\beta}$; the factor $D_{\rm s}^2$ accounts for the
increase of volume with distance. Thus
 $$
\ex{\tau}={\int_0^\infty \!\!\!\tau(D_{\rm s}) \rho(D_{\rm s}) D_{\rm
s}^{2+2\beta}\,\d D_{\rm s}\over
\int_0^\infty \!\!\!\rho(D_{\rm s}) D_{\rm s}^{2+2\beta}\,\d D_{\rm s}}.
\eqno\newe$$

Table 2 lists optical depths $\tau_{-6}\equiv\ex{\tau}/10^{-6}$ in Baade's
window [$(l,b)=(1\deg,-3.9\deg)$], for the three luminosity models that are
shown in Figs.~2--4. In all cases the integrals were evaluated for the
density $\rho(D_{\rm s})=\Upsilon_L\, j_L(D_{\rm s})$ with the scaling of
equations \rhob).  Two values of $\beta$ were considered: $\beta=0$,
appropriate for stars that can be detected at any distance along the
line-of-sight, such as the clump giant stars, and $\beta=-1$, which is more
appropriate for main-sequence stars (Kiraga \& Paczynski 1994, Zhao, Rich 
\& Spergel 1996, Zhao \& Mao 1996).  Figure 6 shows optical depths along the
minor axis $l=0$. The differences between models are everywhere small and
can be quantified in first order by assigning a multiplicative constant to
each model. The difference between the $\beta=0$ and $\beta=-1$ cases is of
order $35\%$ in Baade's window.  Figure 7 shows an optical depth map for one
of the models.  Maps for other models have similar structure.

In comparing these results with observations note that the
self-lensing contribution of the Galactic disk is automatically taken
into account. If we restrict the source stars to be bulge stars,
the optical depth increases by of order $25\%$. For a bulge star at
$8\kpc$ distance in Baade's window, foreground lenses with galactocentric
distances $0-2\kpc$, $2-3\kpc$, $3-4\kpc$, $4-5\kpc$ and $5-8\kpc$
contribute $\tau_{-6}=0.42, 0.19, 0.15, 0.12, 0.21$ to the total
optical depth $\tau_{-6}=1.1$.

The main results from Table 2 and Figs.~6-7 are the following:

(i) The predicted values of $\tau$ are almost independent of the
orientation $\phi_0$ of the bar and of the initial model from which
the iterations were started.

(ii) For $\beta=-1$, the predicted optical depth in Baade's window is
$\tau_{-6}=0.8-0.9$; for comparison, the inferred OGLE and MACHO
values for main-sequence stars are $\tau_{-6}=3.3\pm 2.4$ ($2\sigma$)
from 9 events (Udalski \etal\ 1994), and $\tau_{-6}=1.9\pm0.8$
($2\sigma$) from 41 events (Alcock \etal\ 1995) when the correction
for the disk's contribution is removed.  The predicted optical depth
at the mean $(l,b)=(2.7\deg,-4.08\deg)$ of the MACHO fields is $\simeq
15\%$ lower than in Baade's window. Both quoted measurements are
consistent with our prediction only at the $2-2.5\sigma$ level.

(iii) For 13 clump giants Alcock \etal\ (1995) find
$\tau_{-6}=3.9^{+3.6}_{-2.4}$ ($2\sigma$). These values are averages
over $\sim 12$ square degrees centered at
$(l,b)=(2.55\deg,-3.64\deg)$.  Our predicted optical depth at this
position is $\simeq7\%$ higher than the result for Baade's window.
Thus again our estimate is consistent with the Alcock \etal\ result
only at the $2\sigma$ level.

A more accurate comparison with the observations would involve
averaging the optical depth in Fig.~7 over the various observed
fields, taking into account the number of stars monitored in each
field, the total time monitored per field, etc.

\section{Discussion and Conclusions}

We have shown that models of the Galactic bulge/bar and inner disk
which (i) agree with the COBE/DIRBE $L$-band photometric data, and (ii)
have a constant $L$-band mass-to-light ratio $\Upsilon_L$, fail to
reproduce the reported microlensing optical depths to Baade's window
by a factor of $2-3$ -- the predicted optical depths lie $2-2.5\sigma$ below
the measured values. In this
Section, we summarize the steps that lead to this conclusion and
discuss possible ways out of this discrepancy.

\subsection{The Galactic bar is short, flat, not very elongated,
and not very close to end-on}

The asymmetry maps of simple analytical bars are dominated by twin
peaks.  The location of these peaks varies with the orientation,
elongation and axis ratios of the bar in a way that is readily
understood. In particular, fatter bars have wider peaks, and
lengthening a bar moves its peaks towards higher $l$ and $|b|$. A near
end-on bar can have its peaks at similar locations to those of the
Galaxy, but then its peaks are elongated differently from the
Galaxy's.

Lucy--Richardson iterations that start from one of these simple bars
considerably improve the fit of the model's asymmetry map to that of
the Galaxy by deforming the bar into an approximately standard
shape. As was discussed in Paper III, this has axis ratios
$1{:}0.6{:}0.4$.  The length of the three-dimensional part of this
bar/bulge is around $2 \kpc$. The part of the bar that extends to
$\simeq 3-3.5\kpc$ is strongly flattened to the plane.

If the Sun--centre line is inclined at $\phi_0\simeq25\deg$ to the
long axis of the bar, this standard bar provides a good fit to the
asymmetry map of the COBE data. But if the bar is oriented much more
nearly end-on, it provides an unacceptable fit to the observed
asymmetry map, mainly because the model peaks have the wrong shape.

Moreover, independently of the seriousness with which one views the
failure of near end-on models to fit the observed asymmetry map, there
can be little doubt as to the {\it shape\/} of the near-IR luminosity
density in the inner Galaxy, because this depends so insensitively on
$\phi_0$.

\subsection{Bulge mass and microlensing optical depth for constant
$\Upsilon_L$}

If we assume that the mass-to-light ratio $\Upsilon_L$ is some fixed
number, we can determine this number by comparing simulated
terminal--velocity plots with observed ones. We feel some confidence
in this calibration because it yields an excellent fit to the Galactic
terminal velocity curve (Fig.~5) and because the simulated $(l,v)$
plots are similar to the observed $(l,v)$ diagrams in that they show
features such as the $3\kpc$-arm (Englmaier \& Gerhard 1997).

Given the assumption of constant mass-to-light ratio $\Upsilon_L$, 
the mass distribution of the Galactic bar and inner disk is
rather precisely constrained, and with it the optical depth to
microlensing $\tau$. Nearly independent of the bar angle $\phi_0$,
$\tau_{-6} \equiv \tau / 10^{-6} = 0.8 - 0.9$ for main sequence
sources and $\tau_{-6} = 1.2-1.3$ for clump giants, where the range of
values given reflects the uncertainty due to differences between
various allowed bar models. For a bulge star at $8\kpc$ distance in
Baade's window, foreground `bulge' lenses with galactocentric
distances $0-3\kpc$ contribute $\simeq 55 \%$ to the total optical
depth.  These numerical values are for a tangential velocity of the
LSR of $v_0=200\kms$; if $v_0=220\kms$, the terminal velocity curve is
not fit as well, but for this case the inferred mass-to-light ratio
and optical depths are $10\%$ larger.

These results are inconsistent with the measured optical depths at the
$2\sigma$ level, and are in conflict with the optical depths predicted
by some earlier photometric bar models -- for Baade's window Zhao, Rich
\& Spergel (1996) predict $\tau_{-6}=1.1$ for the bar only, while
Zhao \& Mao (1996) predict optical depths up to $\tau_{-6}=3$.

There appear to be two reasons why our predictions lie below those of
other photometric models:

(i) As discussed above, our model luminosity distribution differs from
those used by other authors in three important respects: it is more
centrally concentrated, more strongly flattened towards the plane and
less elongated. These differences reflect the facts (i) that as a
non-parametric model it fits that data better than do, for example,
the widely used Dwek et al (1995) parametric models, and (ii) it is
based on a much more sophisticated model of dust absorption than the
foreground-screen model of Arendt et al (1994).
The central concentration, the large flattening and the moderate
elongation of our model all conspire to diminish, for a given bulge
mass, the predicted value of $\tau$ in Baade's window because this
field lies $4\deg$ from the plane.

(ii) We determine the mass-to-light ratio $\Upsilon_L$ from gas
velocities rather than from stellar kinematics. In our view gas
velocities provide the securer normalization because it is easier to
model the relevant gas dynamics than the required stellar dynamics.
In particular, it is at present clear neither what the line-of-sight
distribution of any of the observed stellar samples is, nor how
anisotropic the velocity ellipsoids of these stars are at various
points along the line of sight (e.g. Sadler, Rich \& Terndrup
1996). By contrast, the gas streamlines beyond $R\gta 3\kpc$ deviate
from circular orbits by only about $5\%$.

 These differences in the way that we have constructed our bulge model
lead to our bulge mass being relatively small. Integrating over the
total NIR luminosity distribution, multiplying by the normalization of
eqs.~\rhob), and subtracting the mass of the {\it initial\/} disk
models used for the iterations in Paper III yields a bulge mass of
$M_{\rm b}\simeq 7.2-8.6\times 10^9 \msun$ inside $2.4\kpc$. By
contrast, Zhao, Rich \& Spergel (1996) derive $M_{\rm b}\simeq2\times
10^{10} \msun$, while Zhao \& Mao (1996) assume $1.8\times 10^{10} \le
M_{\rm b}/\msun\le 2.8\times 10^{10}$. These masses are more in accord
with the total bulge plus disk mass in our model, which is $M_{\rm
tot}(<2.4\kpc) = 1.9\times 10^{10} \msun$.

\subsection{How to resolve the discrepancy?}

We believe that our models of the inner Galaxy are the most carefully
constructed models available, and that the lensing optical depths that
they predict should be treated with respect. Their ability to give a
good qualitative account of observed $(l,v)$ diagrams (Englmaier \&
Gerhard 1997) gives one considerable confidence in their fundamental
correctness. We trust that this confidence will soon be tested by an
examination of their ability to account equally well for the data on
stellar velocity dispersions in the inner Galaxy -- complete dynamical
models are in an advanced state of preparation. But even now it is
puzzling that there is such significant conflict between the optical
depth obtained from the microlensing experiments and that obtained
from our modelling of the COBE and HI terminal velocity data.  Several
possible causes of this discrepancy stand out.

The first is that the corrections of Spergel et al (1996) for dust
absorption are incorrect. As we have emphasized, the non-axisymmetric
shape of the central luminosity distribution depends upon asymmetries
of $\lta 0.4$ magnitudes in the adopted dust-free brightness
distribution. Consequently, small errors in the absorption corrections
that have been applied to the COBE data could have significant effects
on the final model, especially in the vicinity of the Galactic
plane. However, against this possibility one must count the rigorous
checks which Spergel et al showed their corrections could pass. The
remaining scatter in the dereddened $K-L$ colours is only 0.076 mag,
which they take as a reasonable estimate of the uncertainty in
$L$. Any errors in the absorption corrections are likely to be
significantly smaller in Baade's window than in the Galactic disk at
lower latitudes.  Notice also that any clumpiness of the dust
distribution is automatically taken into account in the scaling to the
NIR reddening.  Finally, the experiment reported below shows that if
the mass in the Galactic plane has been overestimated, the indirect
effects on the optical depth cannot be very large.

Another possible source of error might be the assumption of eight-fold
(triaxial) symmetry assumed in the deprojection of the bulge. If the
main part of the bulge were orientated more end-on than the part that
causes the most prominent features in the difference maps, the optical
depths obtained from our models might underestimate the real
optical depths. There is a limit to the amount by which we can stretch
bulge light along the line-of-sight, however: the bulge/bar proper
must lie within its corotation radius, which from the gas dynamics we
know to be $\sim3\kpc$. Therefore, we can obtain an upper limit on
the magnitude of the effect of rearranging part of the bulge in the
following way. We redistribute the surface density in Baade's window
that arises from the density at line-of-sight distances in the range
$[-3\kpc,3\kpc]$ from the center ($75\%$ of the total), homogeneously
in that distance range. This increases the number of sources at the
far end and the number of lenses at the near end of the bar, so that
the total optical depth increases by $20\%$ (for Model 2). This is a
substantial overestimate of the effect of redistributing some of the
bulge light, because {\it any} bulge, whatever its orientation, will
be inhomogeneous.  Thus the error introduced by our assumption of
eight-fold symmetry must be below $10\%$.

The third possibility is that searches for microlensing events are
more efficient than their practitioners estimate. It is obviously hard
for us to comment on this possibility. Blending effects have been
suggested as a possible cause of overestimating the detection
efficiencies (Alard 1996) at perhaps a $\sim 30\%$ level for main sequence
stars.  However, these effects are expected to be less important for
clump giants.

The final possibility is that the $L$-band mass-to-light ratio varies
significantly with position in the inner Galaxy. Specifically, if
$\Upsilon_L$ were higher above the plane than in the plane, the map of
optical depth shown in Fig.~7 would have too steep a gradient away
from the plane. This might be the case if the $L$-band light
distribution had a significant contribution from young supergiant
stars unrelated to the bulk of the old stellar population, or if there
were significant contribution to the $L$-band light from PAH $3.3 \micron$
or dust emission, or if the distribution of stellar mass were more
concentrated to the plane than the distribution of gravitating mass.

We have tested this by considering mass models in which $\Upsilon_L$
varies as $\Upsilon_L= \Upsilon_{L0}[1 + \tanh(|z|/150\pc)]$.  This
doubles the mass-to-light ratio at large $|z|$ with respect to the
value $\Upsilon_{L0}$ in the plane. On recomputing the potential and
choosing $\Upsilon_{L0}$ so as to keep constant the circular speed at
$3\kpc$, we find that $\Upsilon_{L0}$ is $33\%$ lower than the value
of $\Upsilon_L$ that is given in eq.~(3a). Thus the constraints from
the terminal-velocity curve ensure that the mass of the upper bulge
increases only by a factor of $1.34$ when we assume that $\Upsilon$ is
there twice as great as in the plane. The resulting values for the
optical depth in Baade's window, $\tau_{-6}=1.08$ for $\beta=-1$ main
sequence stars and $\tau_{-6}=1.5$ for $\beta=0$ clump giants, still
lie well below the measured values.

Similarly, even models that include `dark disks' are unlikely to give much
higher optical depths to Baade's window than those calculated here,
for the following reason. The good fit of our model terminal velocity
curve to the observed terminal velocities in Fig.~5 argues that the
{\sl radial} distribution of mass in the model is approximately
correct. Thus the radial mass distribution of any additional component
has to be similar to that of the NIR light, and our derived
mass-to-light ratio automatically includes its contribution. The main
remaining freedom is to shift mass out of the Galactic plane while
keeping the integrated surface density constant. This would be most
effective about half-way between the Sun and the Galactic centre,
where the NIR disk has a vertical exponential scale-length of $h_z
\sim 150\pc$ (Paper III, Fig.~9).  The line-of-sight to Baade's window
passes about $\sim 300\pc$ above the plane at galactocentric radius
$4\kpc$. If the vertical distribution of mass is also exponential, but
with a longer scale-length and correspondingly reduced mass density in
the plane, then the density at height $300\pc$ can be increased by up
to a factor $1.36$. This optimal value occurs for $h_z=300\pc$.
Since thickening the disk reduces the contribution from lenses near
the Sun and has little effect in the bulge, we conclude that by
thickening the mass distribution one cannot increase the optical depth
to Baade's window by more than $\sim 20\%$.

\subsection{Conclusion}

Constant mass-to-light ratio models of the inner Galaxy,
which are consistent with the COBE/DIRBE near-infrared photometry and
the HI and CO terminal velocity curve, result in low optical depths for
bulge microlensing, almost independent of the orientation of the bar:
$\tau_{-6} \equiv \tau / 10^{-6} \simeq 0.9$ for main sequence
sources, and $\tau_{-6} \simeq 1.3$ for clump giants. These values are
inconsistent at the $2-2.5\sigma$ level with the optical depths
inferred from microlensing experiments. We have discussed several
possible uncertainties, including possible variations of the near-infrared
mass-to-light ratio, but none of them appears to be large enough to explain
the discrepancy.

\section*{Acknowledgments} We thank Butler Burton for sending
unpublished HI terminal velocity data.

\bye

%% file: mn.tex
%
%
%
%

\catcode `\@=11 

\def\@version{1.4}
\def\@verdate{22nd Feb 1994}

%
%
%
%


\newif\ifprod@font

\ifx\@typeface\undefined
  \def\@typeface{Comp. Modern}\prod@fontfalse
\else
  \prod@fonttrue 
\fi

\def\newfam{\alloc@8\fam\chardef\sixt@@n} 

\ifprod@font
\font\fiverm=mtr10 at 5pt
\font\fivebf=mtbx10 at 5pt
\font\fiveit=mtti10 at 5pt
\font\fivesl=mtsl10 at 5pt
\font\fivett=mttt10 at 5pt     \hyphenchar\fivett=-1
\font\fivecsc=mtcsc10 at 5pt
\font\fivesf=mtss10 at 5pt
\font\fivei=mtmi10 at 5pt      \skewchar\fivei='177
\font\fivemib=mtmib10 at 5pt   \skewchar\fivemib='177
\font\fivesy=mtsy10 at 5pt     \skewchar\fivesy='60
\font\fivesyb=mtbsy10 at 5pt   \skewchar\fivesyb='60

\font\sixrm=mtr10 at 6pt
\font\sixbf=mtbx10 at 6pt
\font\sixit=mtti10 at 6pt
\font\sixsl=mtsl10 at 6pt
\font\sixtt=mttt10 at 6pt      \hyphenchar\sixtt=-1
\font\sixcsc=mtcsc10 at 6pt
\font\sixsf=mtss10 at 6pt
\font\sixi=mtmi10 at 6pt       \skewchar\sixi='177
\font\sixmib=mtmib10 at 6pt    \skewchar\sixmib='177
\font\sixsy=mtsy10 at 6pt      \skewchar\sixsy='60
\font\sixsyb=mtbsy10 at 6pt    \skewchar\sixsyb='60

\font\sevenrm=mtr10 at 7pt
\font\sevenbf=mtbx10 at 7pt
\font\sevenit=mtti10 at 7pt
\font\sevensl=mtsl10 at 7pt
\font\seventt=mttt10 at 7pt     \hyphenchar\seventt=-1
\font\sevencsc=mtcsc10 at 7pt
\font\sevensf=mtss10 at 7pt
\font\seveni=mtmi10 at 7pt      \skewchar\seveni='177
\font\sevenmib=mtmib10 at 7pt   \skewchar\sevenmib='177
\font\sevensy=mtsy10 at 7pt     \skewchar\sevensy='60
\font\sevensyb=mtbsy10 at 7pt   \skewchar\sevensyb='60

\font\eightrm=mtr10 at 8pt
\font\eightbf=mtbx10 at 8pt
\font\eightit=mtti10 at 8pt
\font\eighti=mtmi10 at 8pt      \skewchar\eighti='177
\font\eightmib=mtmib10 at 8pt   \skewchar\eightmib='177
\font\eightsy=mtsy10 at 8pt     \skewchar\eightsy='60
\font\eightsyb=mtbsy10 at 8pt   \skewchar\eightsyb='60
\font\eightsl=mtsl10 at 8pt
\font\eighttt=mttt10 at 8pt     \hyphenchar\eighttt=-1
\font\eightcsc=mtcsc10 at 8pt
\font\eightsf=mtss10 at 8pt

\font\ninerm=mtr10 at 9pt
\font\ninebf=mtbx10 at 9pt
\font\nineit=mtti10 at 9pt
\font\ninei=mtmi10 at 9pt      \skewchar\ninei='177
\font\ninemib=mtmib10 at 9pt   \skewchar\ninemib='177
\font\ninesy=mtsy10 at 9pt     \skewchar\ninesy='60
\font\ninesyb=mtbsy10 at 9pt   \skewchar\ninesyb='60
\font\ninesl=mtsl10 at 9pt
\font\ninett=mttt10 at 9pt     \hyphenchar\ninett=-1
\font\ninecsc=mtcsc10 at 9pt
\font\ninesf=mtss10 at 9pt

\font\tenrm=mtr10
\font\tenbf=mtbx10
\font\tenit=mtti10
\font\teni=mtmi10		\skewchar\teni='177
\font\tenmib=mtmib10	\skewchar\tenmib='177
\font\tensy=mtsy10		\skewchar\tensy='60
\font\tensyb=mtbsy10	\skewchar\tensyb='60
\font\tenex=cmex10
\font\tensl=mtsl10
\font\tentt=mttt10		\hyphenchar\tentt=-1
\font\tencsc=mtcsc10
\font\tensf=mtss10

\font\elevenrm=mtr10 at 11pt
\font\elevenbf=mtbx10 at 11pt
\font\elevenit=mtti10 at 11pt
\font\eleveni=mtmi10 at 11pt      \skewchar\eleveni='177
\font\elevenmib=mtmib10 at 11pt   \skewchar\elevenmib='177
\font\elevensy=mtsy10 at 11pt     \skewchar\elevensy='60
\font\elevensyb=mtbsy10 at 11pt   \skewchar\elevensyb='60
\font\elevensl=mtsl10 at 11pt
\font\eleventt=mttt10 at 11pt     \hyphenchar\eleventt=-1
\font\elevencsc=mtcsc10 at 11pt
\font\elevensf=mtss10 at 11pt

\font\twelverm=mtr10 at 12pt
\font\twelvebf=mtbx10 at 12pt
\font\twelveit=mtti10 at 12pt
\font\twelvesl=mtsl10 at 12pt
\font\twelvett=mttt10 at 12pt     \hyphenchar\twelvett=-1
\font\twelvecsc=mtcsc10 at 12pt
\font\twelvesf=mtss10 at 12pt
\font\twelvei=mtmi10 at 12pt      \skewchar\twelvei='177
\font\twelvemib=mtmib10 at 12pt   \skewchar\twelvemib='177
\font\twelvesy=mtsy10 at 12pt     \skewchar\twelvesy='60
\font\twelvesyb=mtbsy10 at 12pt   \skewchar\twelvesyb='60

\font\fourteenrm=mtr10 at 14pt
\font\fourteenbf=mtbx10 at 14pt
\font\fourteenit=mtti10 at 14pt
\font\fourteeni=mtmi10 at 14pt      \skewchar\fourteeni='177
\font\fourteenmib=mtmib10 at 14pt   \skewchar\fourteenmib='177
\font\fourteensy=mtsy10 at 14pt     \skewchar\fourteensy='60
\font\fourteensyb=mtbsy10 at 14pt   \skewchar\fourteensyb='60
\font\fourteensl=mtsl10 at 14pt
\font\fourteentt=mttt10 at 14pt     \hyphenchar\fourteentt=-1
\font\fourteencsc=mtcsc10 at 14pt
\font\fourteensf=mtss10 at 14pt

\font\seventeenrm=mtr10 at 17pt
\font\seventeenbf=mtbx10 at 17pt
\font\seventeenit=mtti10 at 17pt
\font\seventeeni=mtmi10 at 17pt      \skewchar\seventeeni='177
\font\seventeenmib=mtmib10 at 17pt   \skewchar\seventeenmib='177
\font\seventeensy=mtsy10 at 17pt     \skewchar\seventeensy='60
\font\seventeensyb=mtbsy10 at 17pt   \skewchar\seventeensyb='60
\font\seventeensl=mtsl10 at 17pt
\font\seventeentt=mttt10 at 17pt     \hyphenchar\seventeentt=-1
\font\seventeencsc=mtcsc10 at 17pt
\font\seventeensf=mtss10 at 17pt


\newfam\xmfam
\newfam\ymfam

\font\fivexm=mtxm10 at 5pt
\font\sixxm=mtxm10 at 6pt
\font\sevenxm=mtxm10 at 7pt
\font\eightxm=mtxm10 at 8pt
\font\ninexm=mtxm10 at 9pt
\font\tenxm=mtxm10
\font\elevenxm=mtxm10 at 11pt
\font\twelvexm=mtxm10 at 12pt
\font\fourteenxm=mtxm10 at 14pt
\font\seventeenxm=mtxm10 at 17pt

\font\fiveym=mtym10 at 5pt
\font\sixym=mtym10 at 6pt
\font\sevenym=mtym10 at 7pt
\font\eightym=mtym10 at 8pt
\font\nineym=mtym10 at 9pt
\font\tenym=mtym10
\font\elevenym=mtym10 at 11pt
\font\twelveym=mtym10 at 12pt
\font\fourteenym=mtym10 at 14pt
\font\seventeenym=mtym10 at 17pt
\else
\font\fiverm=cmr5
\font\fivei=cmmi5             \skewchar\fivei='177
\font\fivemib=cmmib10 at 5pt  \skewchar\fivemib='177
\font\fivesy=cmsy5            \skewchar\fivesy='60
\font\fivesyb=cmbsy10 at 5pt  \skewchar\fivesyb='60
\font\fivebf=cmbx5

\font\sixrm=cmr6
\font\sixi=cmmi6             \skewchar\sixi='177
\font\sixmib=cmmib10 at 6pt  \skewchar\sixmib='177
\font\sixsy=cmsy6            \skewchar\sixsy='60
\font\sixsyb=cmbsy10 at 6pt  \skewchar\sixsyb='60
\font\sixbf=cmbx6

\font\sevenrm=cmr7
\font\seveni=cmmi7             \skewchar\seveni='177
\font\sevenmib=cmmib10 at 7pt  \skewchar\sevenmib='177
\font\sevensy=cmsy7            \skewchar\sevensy='60
\font\sevensyb=cmbsy10 at 7pt  \skewchar\sevensyb='60
\font\sevenbf=cmbx7

\font\eightrm=cmr8
\font\eightbf=cmbx8
\font\eightit=cmti8
\font\eighti=cmmi8			\skewchar\eighti='177
\font\eightmib=cmmib10 at 8pt	\skewchar\eightmib='177
\font\eightsy=cmsy8			\skewchar\eightsy='60
\font\eightsyb=cmbsy10 at 8pt	\skewchar\eightsyb='60
\font\eightsl=cmsl8
\font\eighttt=cmtt8			\hyphenchar\eighttt=-1
\font\eightcsc=cmcsc10 at 8pt
\font\eightsf=cmss8

\font\ninerm=cmr9
\font\ninebf=cmbx9
\font\nineit=cmti9
\font\ninei=cmmi9			\skewchar\ninei='177
\font\ninemib=cmmib10 at 9pt	\skewchar\ninemib='177
\font\ninesy=cmsy9			\skewchar\ninesy='60
\font\ninesyb=cmbsy10 at 9pt	\skewchar\ninesyb='60
\font\ninesl=cmsl9
\font\ninett=cmtt9			\hyphenchar\ninett=-1
\font\ninecsc=cmcsc10 at 9pt
\font\ninesf=cmss9

\font\tenrm=cmr10
\font\tenbf=cmbx10
\font\tenit=cmti10
\font\teni=cmmi10		\skewchar\teni='177
\font\tenmib=cmmib10	\skewchar\tenmib='177
\font\tensy=cmsy10		\skewchar\tensy='60
\font\tensyb=cmbsy10	\skewchar\tensyb='60
\font\tenex=cmex10
\font\tensl=cmsl10
\font\tentt=cmtt10		\hyphenchar\tentt=-1
\font\tencsc=cmcsc10
\font\tensf=cmss10

\font\elevenrm=cmr10 scaled \magstephalf
\font\elevenbf=cmbx10 scaled \magstephalf
\font\elevenit=cmti10 scaled \magstephalf
\font\eleveni=cmmi10 scaled \magstephalf	\skewchar\eleveni='177
\font\elevenmib=cmmib10 scaled \magstephalf	\skewchar\elevenmib='177
\font\elevensy=cmsy10 scaled \magstephalf	\skewchar\elevensy='60
\font\elevensyb=cmbsy10 scaled \magstephalf	\skewchar\elevensyb='60
\font\elevensl=cmsl10 scaled \magstephalf
\font\eleventt=cmtt10 scaled \magstephalf	\hyphenchar\eleventt=-1
\font\elevencsc=cmcsc10 scaled \magstephalf
\font\elevensf=cmss10 scaled \magstephalf

\font\twelverm=cmr10 scaled \magstep1
\font\twelvebf=cmbx10 scaled \magstep1
\font\twelvei=cmmi10 scaled \magstep1      \skewchar\twelvei='177
\font\twelvemib=cmmib10 scaled \magstep1   \skewchar\twelvemib='177
\font\twelvesy=cmsy10 scaled \magstep1     \skewchar\twelvesy='60
\font\twelvesyb=cmbsy10 scaled \magstep1   \skewchar\twelvesyb='60

\font\fourteenrm=cmr10 scaled \magstep2
\font\fourteenbf=cmbx10 scaled \magstep2
\font\fourteenit=cmti10 scaled \magstep2
\font\fourteeni=cmmi10 scaled \magstep2		\skewchar\fourteeni='177
\font\fourteenmib=cmmib10 scaled \magstep2	\skewchar\fourteenmib='177
\font\fourteensy=cmsy10 scaled \magstep2	\skewchar\fourteensy='60
\font\fourteensyb=cmbsy10 scaled \magstep2	\skewchar\fourteensyb='60
\font\fourteensl=cmsl10 scaled \magstep2
\font\fourteentt=cmtt10 scaled \magstep2	\hyphenchar\fourteentt=-1
\font\fourteencsc=cmcsc10 scaled \magstep2
\font\fourteensf=cmss10 scaled \magstep2

\font\seventeenrm=cmr10 scaled \magstep3
\font\seventeenbf=cmbx10 scaled \magstep3
\font\seventeenit=cmti10 scaled \magstep3
\font\seventeeni=cmmi10 scaled \magstep3	\skewchar\seventeeni='177
\font\seventeenmib=cmmib10 scaled \magstep3	\skewchar\seventeenmib='177
\font\seventeensy=cmsy10 scaled \magstep3	\skewchar\seventeensy='60
\font\seventeensyb=cmbsy10 scaled \magstep3	\skewchar\seventeensyb='60
\font\seventeensl=cmsl10 scaled \magstep3
\font\seventeentt=cmtt10 scaled \magstep3	\hyphenchar\seventeentt=-1
\font\seventeencsc=cmcsc10 scaled \magstep3
\font\seventeensf=cmss10 scaled \magstep3
\fi

\def\hexnumber#1{\ifcase#1 0\or1\or2\or3\or4\or5\or6\or7\or8\or9\or
  A\or B\or C\or D\or E\or F\fi}

\ifprod@font
  \edef\@xm{\hexnumber\xmfam}
  \edef\@ym{\hexnumber\ymfam}
\fi

\def\makestrut{%
  \setbox\strutbox=\hbox{%
    \vrule height.7\baselineskip depth.3\baselineskip width \z@}%
}

\def\baselinestretch{1}
\newskip\tmp@bls

\def\b@ls#1{
  \tmp@bls=#1\relax
  \baselineskip=#1\relax\makestrut
  \normalbaselineskip=\baselinestretch\tmp@bls
  \normalbaselines
}

\def\nostb@ls#1{
  \normalbaselineskip=#1\relax
  \normalbaselines
  \makestrut
}

%

\newfam\mibfam 
\newfam\sybfam 
\newfam\scfam  
\newfam\sffam  

\def\mit{\fam\@ne}

\def\cal{\fam\tw@}

\def\em{\ifdim\fontdimen1\font>\z@ \rm\else\it\fi}

\textfont3=\tenex
\scriptfont3=\tenex
\scriptscriptfont3=\tenex

\setbox0=\hbox{\tenex B} \p@renwd=\wd0 

\def\eightpoint{
  \def\rm{\fam0\eightrm}%
  \textfont0=\eightrm \scriptfont0=\sixrm \scriptscriptfont0=\fiverm%
  \textfont1=\eighti  \scriptfont1=\sixi  \scriptscriptfont1=\fivei%
  \textfont2=\eightsy \scriptfont2=\sixsy \scriptscriptfont2=\fivesy%
  \textfont\itfam=\eightit\def\it{\fam\itfam\eightit}%
  \ifprod@font
    \scriptfont\itfam=\sixit
      \scriptscriptfont\itfam=\fiveit
  \else
    \scriptfont\itfam=\eightit
      \scriptscriptfont\itfam=\eightit
  \fi
  \textfont\bffam=\eightbf%
    \scriptfont\bffam=\sixbf%
      \scriptscriptfont\bffam=\fivebf%
  \def\bf{\fam\bffam\eightbf}%
  \textfont\slfam=\eightsl\def\sl{\fam\slfam\eightsl}%
  \ifprod@font
    \scriptfont\slfam=\sixsl
      \scriptscriptfont\slfam=\fivesl
  \else
    \scriptfont\slfam=\eightsl
      \scriptscriptfont\slfam=\eightsl
  \fi
  \textfont\ttfam=\eighttt\def\tt{\fam\ttfam\eighttt}%
  \ifprod@font
    \scriptfont\ttfam=\sixtt
      \scriptscriptfont\ttfam=\fivett
  \else
    \scriptfont\ttfam=\eighttt
      \scriptscriptfont\ttfam=\eighttt
  \fi
  \textfont\scfam=\eightcsc\def\sc{\fam\scfam\eightcsc}%
  \ifprod@font
    \scriptfont\scfam=\sixcsc
      \scriptscriptfont\scfam=\fivecsc
  \else
    \scriptfont\scfam=\eightcsc
      \scriptscriptfont\scfam=\eightcsc
  \fi
  \textfont\sffam=\eightsf\def\sf{\fam\sffam\eightsf}%
  \ifprod@font
    \scriptfont\sffam=\sixsf
      \scriptscriptfont\sffam=\fivesf
  \else
    \scriptfont\sffam=\eightsf
      \scriptscriptfont\sffam=\eightsf
  \fi
  \textfont\mibfam=\eightmib
    \scriptfont\mibfam=\sixmib
      \scriptscriptfont\mibfam=\fivemib
  \textfont\sybfam=\eightsyb
    \scriptfont\sybfam=\sixsyb
      \scriptscriptfont\sybfam=\fivesyb
  \ifprod@font
    \textfont\xmfam=\eightxm
      \scriptfont\xmfam=\sixxm
        \scriptscriptfont\xmfam=\fivexm
    \textfont\ymfam=\eightym
      \scriptfont\ymfam=\sixym
        \scriptscriptfont\ymfam=\fiveym
  \fi
  \def\oldstyle{\fam\@ne\eighti}%
  \def\boldstyle{\fam\mibfam\eightmib}%
  \b@ls{10pt}\rm%
}

\def\ninepoint{
  \def\rm{\fam0\ninerm}%
  \textfont0=\ninerm \scriptfont0=\sixrm \scriptscriptfont0=\fiverm%
  \textfont1=\ninei  \scriptfont1=\sixi  \scriptscriptfont1=\fivei%
  \textfont2=\ninesy \scriptfont2=\sixsy \scriptscriptfont2=\fivesy%
  \textfont\itfam=\nineit\def\it{\fam\itfam\nineit}%
  \ifprod@font
    \scriptfont\itfam=\sixit
      \scriptscriptfont\itfam=\fiveit
  \else
    \scriptfont\itfam=\nineit
      \scriptscriptfont\itfam=\nineit
  \fi
  \textfont\bffam=\ninebf%
    \scriptfont\bffam=\sixbf%
      \scriptscriptfont\bffam=\fivebf%
  \def\bf{\fam\bffam\ninebf}%
  \textfont\slfam=\ninesl\def\sl{\fam\slfam\ninesl}%
  \ifprod@font
    \scriptfont\slfam=\sixsl
      \scriptscriptfont\slfam=\fivesl
  \else
    \scriptfont\slfam=\ninesl
      \scriptscriptfont\slfam=\ninesl
  \fi
  \textfont\ttfam=\ninett\def\tt{\fam\ttfam\ninett}%
  \ifprod@font
    \scriptfont\ttfam=\sixtt
      \scriptscriptfont\ttfam=\fivett
  \else
    \scriptfont\ttfam=\ninett
      \scriptscriptfont\ttfam=\ninett
  \fi
  \textfont\scfam=\ninecsc\def\sc{\fam\scfam\ninecsc}%
  \ifprod@font
    \scriptfont\scfam=\sixcsc
      \scriptscriptfont\scfam=\fivecsc
  \else
    \scriptfont\scfam=\ninecsc
      \scriptscriptfont\scfam=\ninecsc
  \fi
  \textfont\sffam=\ninesf\def\sf{\fam\sffam\ninesf}%
  \ifprod@font
    \scriptfont\sffam=\sixsf
      \scriptscriptfont\sffam=\fivesf
  \else
    \scriptfont\sffam=\ninesf
      \scriptscriptfont\sffam=\ninesf
  \fi
  \textfont\mibfam=\ninemib
    \scriptfont\mibfam=\sixmib
      \scriptscriptfont\mibfam=\fivemib
  \textfont\sybfam=\ninesyb
    \scriptfont\sybfam=\sixsyb
      \scriptscriptfont\sybfam=\fivesyb
  \ifprod@font
    \textfont\xmfam=\ninexm
      \scriptfont\xmfam=\sixxm
        \scriptscriptfont\xmfam=\fivexm
    \textfont\ymfam=\nineym
      \scriptfont\ymfam=\sixym
        \scriptscriptfont\ymfam=\fiveym
  \fi
  \def\oldstyle{\fam\@ne\ninei}%
  \def\boldstyle{\fam\mibfam\ninemib}%
  \b@ls{\TextLeading plus \Feathering}\rm%
}

\def\tenpoint{
  \def\rm{\fam0\tenrm}%
  \textfont0=\tenrm \scriptfont0=\sevenrm \scriptscriptfont0=\fiverm%
  \textfont1=\teni  \scriptfont1=\seveni  \scriptscriptfont1=\fivei%
  \textfont2=\tensy \scriptfont2=\sevensy \scriptscriptfont2=\fivesy%
  \textfont\itfam=\tenit\def\it{\fam\itfam\tenit}%
  \ifprod@font
    \scriptfont\itfam=\sevenit
      \scriptscriptfont\itfam=\fiveit
  \else
    \scriptfont\itfam=\tenit
      \scriptscriptfont\itfam=\tenit
  \fi
  \textfont\bffam=\tenbf%
    \scriptfont\bffam=\sevenbf%
      \scriptscriptfont\bffam=\fivebf%
  \def\bf{\fam\bffam\tenbf}%
  \textfont\slfam=\tensl\def\sl{\fam\slfam\tensl}%
  \ifprod@font
    \scriptfont\slfam=\sevensl
      \scriptscriptfont\slfam=\fivesl
  \else
    \scriptfont\slfam=\tensl
      \scriptscriptfont\slfam=\tensl
  \fi
  \textfont\ttfam=\tentt\def\tt{\fam\ttfam\tentt}%
  \ifprod@font
    \scriptfont\ttfam=\seventt
      \scriptscriptfont\ttfam=\fivett
  \else
    \scriptfont\ttfam=\tentt
      \scriptscriptfont\ttfam=\tentt
  \fi
  \textfont\scfam=\tencsc\def\sc{\fam\scfam\tencsc}%
  \ifprod@font
    \scriptfont\scfam=\sevencsc
      \scriptscriptfont\scfam=\fivecsc
  \else
    \scriptfont\scfam=\tencsc
      \scriptscriptfont\scfam=\tencsc
  \fi
  \textfont\sffam=\tensf\def\sf{\fam\sffam\tensf}%
  \ifprod@font
    \scriptfont\sffam=\sevensf
      \scriptscriptfont\sffam=\fivesf
  \else
    \scriptfont\sffam=\tensf
      \scriptscriptfont\sffam=\tensf
  \fi
  \textfont\mibfam=\tenmib
    \scriptfont\mibfam=\sevenmib
      \scriptscriptfont\mibfam=\fivemib
  \textfont\sybfam=\tensyb
    \scriptfont\sybfam=\sevensyb
      \scriptscriptfont\sybfam=\fivesyb
  \ifprod@font
    \textfont\xmfam=\tenxm
      \scriptfont\xmfam=\sevenxm
        \scriptscriptfont\xmfam=\fivexm
    \textfont\ymfam=\tenym
      \scriptfont\ymfam=\sevenym
        \scriptscriptfont\ymfam=\fiveym
  \fi
  \def\oldstyle{\fam\@ne\teni}%
  \def\boldstyle{\fam\mibfam\tenmib}%
  \b@ls{11pt}\rm%
}

\def\elevenpoint{
  \def\rm{\fam0\elevenrm}%
  \textfont0=\elevenrm \scriptfont0=\eightrm \scriptscriptfont0=\sixrm%
  \textfont1=\eleveni  \scriptfont1=\eighti  \scriptscriptfont1=\sixi%
  \textfont2=\elevensy \scriptfont2=\eightsy \scriptscriptfont2=\sixsy%
  \textfont\itfam=\elevenit\def\it{\fam\itfam\elevenit}%
  \ifprod@font
    \scriptfont\itfam=\eightit
      \scriptscriptfont\itfam=\sixit
  \else
    \scriptfont\itfam=\elevenit
      \scriptscriptfont\itfam=\elevenit
  \fi
  \textfont\bffam=\elevenbf%
    \scriptfont\bffam=\eightbf%
      \scriptscriptfont\bffam=\sixbf%
  \def\bf{\fam\bffam\elevenbf}%
  \textfont\slfam=\elevensl\def\sl{\fam\slfam\elevensl}%
  \ifprod@font
    \scriptfont\slfam=\eightsl
      \scriptscriptfont\slfam=\sixsl
  \else
    \scriptfont\slfam=\elevensl
      \scriptscriptfont\slfam=\elevensl
  \fi
  \textfont\ttfam=\eleventt\def\tt{\fam\ttfam\eleventt}%
  \ifprod@font
    \scriptfont\ttfam=\eighttt
      \scriptscriptfont\ttfam=\sixtt
  \else
    \scriptfont\ttfam=\eleventt
      \scriptscriptfont\ttfam=\eleventt
  \fi
  \textfont\scfam=\elevencsc\def\sc{\fam\scfam\elevencsc}%
  \ifprod@font
    \scriptfont\scfam=\eightcsc
      \scriptscriptfont\scfam=\sixcsc
  \else
    \scriptfont\scfam=\elevencsc
      \scriptscriptfont\scfam=\elevencsc
  \fi
  \textfont\sffam=\elevensf\def\sf{\fam\sffam\elevensf}%
  \ifprod@font
    \scriptfont\sffam=\eightsf
      \scriptscriptfont\sffam=\sixsf
  \else
    \scriptfont\sffam=\elevensf
      \scriptscriptfont\sffam=\elevensf
  \fi
  \textfont\mibfam=\elevenmib
    \scriptfont\mibfam=\eightmib
      \scriptscriptfont\mibfam=\sixmib
  \textfont\sybfam=\elevensyb
    \scriptfont\sybfam=\eightsyb
      \scriptscriptfont\sybfam=\sixsyb
  \ifprod@font
    \textfont\xmfam=\elevenxm
      \scriptfont\xmfam=\eightxm
       \scriptscriptfont\xmfam=\sixxm
    \textfont\ymfam=\elevenym
      \scriptfont\ymfam=\eightym
        \scriptscriptfont\ymfam=\sixym
   \fi
  \def\oldstyle{\fam\@ne\eleveni}%
  \def\boldstyle{\fam\mibfam\elevenmib}%
  \b@ls{13pt}\rm%
}

\def\fourteenpoint{
  \def\rm{\fam0\fourteenrm}%
  \textfont0\fourteenrm  \scriptfont0\tenrm  \scriptscriptfont0\sevenrm%
  \textfont1\fourteeni   \scriptfont1\teni   \scriptscriptfont1\seveni%
  \textfont2\fourteensy  \scriptfont2\tensy  \scriptscriptfont2\sevensy%
  \textfont\itfam=\fourteenit\def\it{\fam\itfam\fourteenit}%
  \ifprod@font
    \scriptfont\itfam=\tenit
      \scriptscriptfont\itfam=\sevenit
  \else
    \scriptfont\itfam=\fourteenit
      \scriptscriptfont\itfam=\fourteenit
  \fi
  \textfont\bffam=\fourteenbf%
    \scriptfont\bffam=\tenbf%
      \scriptscriptfont\bffam=\sevenbf%
  \def\bf{\fam\bffam\fourteenbf}%
  \textfont\slfam=\fourteensl\def\sl{\fam\slfam\fourteensl}%
  \ifprod@font
    \scriptfont\slfam=\tensl
      \scriptscriptfont\slfam=\sevensl
  \else
    \scriptfont\slfam=\fourteensl
      \scriptscriptfont\slfam=\fourteensl
  \fi
  \textfont\ttfam=\fourteentt\def\tt{\fam\ttfam\fourteentt}%
  \ifprod@font
    \scriptfont\ttfam=\tentt
      \scriptscriptfont\ttfam=\seventt
  \else
    \scriptfont\ttfam=\fourteentt
      \scriptscriptfont\ttfam=\fourteentt
  \fi
  \textfont\scfam=\fourteencsc\def\sc{\fam\scfam\fourteencsc}%
  \ifprod@font
    \scriptfont\scfam=\tencsc
      \scriptscriptfont\scfam=\sevencsc
  \else
    \scriptfont\scfam=\fourteencsc
      \scriptscriptfont\scfam=\fourteencsc
  \fi
  \textfont\sffam=\fourteensf\def\sf{\fam\sffam\fourteensf}%
  \ifprod@font
    \scriptfont\sffam=\tensf
      \scriptscriptfont\sffam=\sevensf
  \else
    \scriptfont\sffam=\fourteensf
      \scriptscriptfont\sffam=\fourteensf
  \fi
  \textfont\mibfam=\fourteenmib
    \scriptfont\mibfam=\tenmib
      \scriptscriptfont\mibfam=\sevenmib
  \textfont\sybfam=\fourteensyb
    \scriptfont\sybfam=\tensyb
      \scriptscriptfont\sybfam=\sevensyb
  \ifprod@font
    \textfont\xmfam=\fourteenxm
      \scriptfont\xmfam=\tenxm
        \scriptscriptfont\xmfam=\sevenxm
   \textfont\ymfam=\fourteenym
      \scriptfont\ymfam=\tenym
        \scriptscriptfont\ymfam=\sevenym
  \fi
  \def\oldstyle{\fam\@ne\fourteeni}%
  \def\boldstyle{\fam\mibfam\fourteenmib}%
  \b@ls{17pt}\rm%
}

\def\seventeenpoint{
  \def\rm{\fam0\seventeenrm}%
  \textfont0\seventeenrm  \scriptfont0\twelverm  \scriptscriptfont0\tenrm%
  \textfont1\seventeeni   \scriptfont1\twelvei   \scriptscriptfont1\teni%
  \textfont2\seventeensy  \scriptfont2\twelvesy  \scriptscriptfont2\tensy%
  \textfont\itfam=\seventeenit\def\it{\fam\itfam\seventeenit}%
  \ifprod@font
    \scriptfont\itfam=\twelveit
      \scriptscriptfont\itfam=\tenit
  \else
    \scriptfont\itfam=\seventeenit
      \scriptscriptfont\itfam=\seventeenit
  \fi
  \textfont\bffam=\seventeenbf%
    \scriptfont\bffam=\twelvebf%
      \scriptscriptfont\bffam=\tenbf%
  \def\bf{\fam\bffam\seventeenbf}%
  \textfont\slfam=\seventeensl\def\sl{\fam\slfam\seventeensl}%
  \ifprod@font
    \scriptfont\slfam=\twelvesl
      \scriptscriptfont\slfam=\tensl
  \else
    \scriptfont\slfam=\seventeensl
      \scriptscriptfont\slfam=\seventeensl
  \fi
  \textfont\ttfam=\seventeentt\def\tt{\fam\ttfam\seventeentt}%
  \ifprod@font
    \scriptfont\ttfam=\twelvett
      \scriptscriptfont\ttfam=\tentt
  \else
    \scriptfont\ttfam=\seventeentt
      \scriptscriptfont\ttfam=\seventeentt
  \fi
  \textfont\scfam=\seventeencsc\def\sc{\fam\scfam\seventeencsc}%
  \ifprod@font
    \scriptfont\scfam=\twelvecsc
      \scriptscriptfont\scfam=\tencsc
  \else
    \scriptfont\scfam=\seventeencsc
      \scriptscriptfont\scfam=\seventeencsc
  \fi
  \textfont\sffam=\seventeensf\def\sf{\fam\sffam\seventeensf}%
  \ifprod@font
    \scriptfont\sffam=\twelvesf
      \scriptscriptfont\sffam=\tensf
  \else
    \scriptfont\sffam=\seventeensf
      \scriptscriptfont\sffam=\seventeensf
  \fi
  \textfont\mibfam=\seventeenmib
    \scriptfont\mibfam=\twelvemib
      \scriptscriptfont\mibfam=\tenmib
  \textfont\sybfam=\seventeensyb
    \scriptfont\sybfam=\twelvesyb
      \scriptscriptfont\sybfam=\tensyb
  \ifprod@font
    \textfont\xmfam=\seventeenxm
      \scriptfont\xmfam=\twelvexm
        \scriptscriptfont\xmfam=\tenxm
    \textfont\ymfam=\seventeenym
      \scriptfont\ymfam=\twelveym
        \scriptscriptfont\ymfam=\tenym
  \fi
  \def\oldstyle{\fam\@ne\seventeeni}%
  \def\boldstyle{\fam\mibfam\seventeenmib}%
  \b@ls{20pt}\rm%
}

\lineskip=1pt      \normallineskip=\lineskip
\lineskiplimit=\z@ \normallineskiplimit=\lineskiplimit



\def\la{\mathrel{\mathchoice {\vcenter{\offinterlineskip\halign{\hfil
$\displaystyle##$\hfil\cr<\cr\sim\cr}}}
{\vcenter{\offinterlineskip\halign{\hfil$\textstyle##$\hfil\cr
<\cr\sim\cr}}}
{\vcenter{\offinterlineskip\halign{\hfil$\scriptstyle##$\hfil\cr
<\cr\sim\cr}}}
{\vcenter{\offinterlineskip\halign{\hfil$\scriptscriptstyle##$\hfil\cr
<\cr\sim\cr}}}}}

\def\ga{\mathrel{\mathchoice {\vcenter{\offinterlineskip\halign{\hfil
$\displaystyle##$\hfil\cr>\cr\sim\cr}}}
{\vcenter{\offinterlineskip\halign{\hfil$\textstyle##$\hfil\cr
>\cr\sim\cr}}}
{\vcenter{\offinterlineskip\halign{\hfil$\scriptstyle##$\hfil\cr
>\cr\sim\cr}}}
{\vcenter{\offinterlineskip\halign{\hfil$\scriptscriptstyle##$\hfil\cr
>\cr\sim\cr}}}}}

\def\getsto{\mathrel{\mathchoice {\vcenter{\offinterlineskip
\halign{\hfil
$\displaystyle##$\hfil\cr\gets\cr\to\cr}}}
{\vcenter{\offinterlineskip\halign{\hfil$\textstyle##$\hfil\cr\gets
\cr\to\cr}}}
{\vcenter{\offinterlineskip\halign{\hfil$\scriptstyle##$\hfil\cr\gets
\cr\to\cr}}}
{\vcenter{\offinterlineskip\halign{\hfil$\scriptscriptstyle##$\hfil\cr
\gets\cr\to\cr}}}}}

\def\lid{\mathrel{\mathchoice {\vcenter{\offinterlineskip\halign{\hfil
$\displaystyle##$\hfil\cr<\cr\noalign{\vskip1.2pt}=\cr}}}
{\vcenter{\offinterlineskip\halign{\hfil$\textstyle##$\hfil\cr<\cr
\noalign{\vskip1.2pt}=\cr}}}
{\vcenter{\offinterlineskip\halign{\hfil$\scriptstyle##$\hfil\cr<\cr
\noalign{\vskip1pt}=\cr}}}
{\vcenter{\offinterlineskip\halign{\hfil$\scriptscriptstyle##$\hfil\cr
<\cr
\noalign{\vskip0.9pt}=\cr}}}}}

\def\gid{\mathrel{\mathchoice {\vcenter{\offinterlineskip\halign{\hfil
$\displaystyle##$\hfil\cr>\cr\noalign{\vskip1.2pt}=\cr}}}
{\vcenter{\offinterlineskip\halign{\hfil$\textstyle##$\hfil\cr>\cr
\noalign{\vskip1.2pt}=\cr}}}
{\vcenter{\offinterlineskip\halign{\hfil$\scriptstyle##$\hfil\cr>\cr
\noalign{\vskip1pt}=\cr}}}
{\vcenter{\offinterlineskip\halign{\hfil$\scriptscriptstyle##$\hfil\cr
>\cr
\noalign{\vskip0.9pt}=\cr}}}}}

\def\grole{\mathrel{\mathchoice {\vcenter{\offinterlineskip\halign{\hfil
$\displaystyle##$\hfil\cr>\cr\noalign{\vskip-1.5pt}<\cr}}}
{\vcenter{\offinterlineskip\halign{\hfil$\textstyle##$\hfil\cr
>\cr\noalign{\vskip-1.5pt}<\cr}}}
{\vcenter{\offinterlineskip\halign{\hfil$\scriptstyle##$\hfil\cr
>\cr\noalign{\vskip-1pt}<\cr}}}
{\vcenter{\offinterlineskip\halign{\hfil$\scriptscriptstyle##$\hfil\cr
>\cr\noalign{\vskip-0.5pt}<\cr}}}}}

\def\leogr{\mathrel{\mathchoice {\vcenter{\offinterlineskip\halign{\hfil
$\displaystyle##$\hfil\cr<\cr\noalign{\vskip-1.5pt}>\cr}}}
{\vcenter{\offinterlineskip\halign{\hfil$\textstyle##$\hfil\cr
<\cr\noalign{\vskip-1.5pt}>\cr}}}
{\vcenter{\offinterlineskip\halign{\hfil$\scriptstyle##$\hfil\cr
<\cr\noalign{\vskip-1pt}>\cr}}}
{\vcenter{\offinterlineskip\halign{\hfil$\scriptscriptstyle##$\hfil\cr
<\cr\noalign{\vskip-0.5pt}>\cr}}}}}

\def\loa{\mathrel{\mathchoice {\vcenter{\offinterlineskip\halign{\hfil
$\displaystyle##$\hfil\cr<\cr\approx\cr}}}
{\vcenter{\offinterlineskip\halign{\hfil$\textstyle##$\hfil\cr
<\cr\approx\cr}}}
{\vcenter{\offinterlineskip\halign{\hfil$\scriptstyle##$\hfil\cr
<\cr\approx\cr}}}
{\vcenter{\offinterlineskip\halign{\hfil$\scriptscriptstyle##$\hfil\cr
<\cr\approx\cr}}}}}

\def\goa{\mathrel{\mathchoice {\vcenter{\offinterlineskip\halign{\hfil
$\displaystyle##$\hfil\cr>\cr\approx\cr}}}
{\vcenter{\offinterlineskip\halign{\hfil$\textstyle##$\hfil\cr
>\cr\approx\cr}}}
{\vcenter{\offinterlineskip\halign{\hfil$\scriptstyle##$\hfil\cr
>\cr\approx\cr}}}
{\vcenter{\offinterlineskip\halign{\hfil$\scriptscriptstyle##$\hfil\cr
>\cr\approx\cr}}}}}

\def\diameter{{\ifmmode\mathchoice
{\ooalign{\hfil\hbox{$\displaystyle/$}\hfil\crcr
{\hbox{$\displaystyle\mathchar"20D$}}}}
{\ooalign{\hfil\hbox{$\textstyle/$}\hfil\crcr
{\hbox{$\textstyle\mathchar"20D$}}}}
{\ooalign{\hfil\hbox{$\scriptstyle/$}\hfil\crcr
{\hbox{$\scriptstyle\mathchar"20D$}}}}
{\ooalign{\hfil\hbox{$\scriptscriptstyle/$}\hfil\crcr
{\hbox{$\scriptscriptstyle\mathchar"20D$}}}}
\else{\ooalign{\hfil/\hfil\crcr\mathhexbox20D}}%
\fi}}

\def\sq{\ifmmode\squareforqed\else{\unskip\nobreak\hfil
\penalty50\hskip1em\null\nobreak\hfil\squareforqed
\parfillskip=0pt\finalhyphendemerits=0\endgraf}\fi}
\def\squareforqed{\hbox{\rlap{$\sqcap$}$\sqcup$}}


\def\bbbc{{\mathchoice {\setbox0=\hbox{$\displaystyle\rm C$}\hbox{\hbox
to0pt{\kern0.4\wd0\vrule height0.9\ht0\hss}\box0}}
{\setbox0=\hbox{$\textstyle\rm C$}\hbox{\hbox
to0pt{\kern0.4\wd0\vrule height0.9\ht0\hss}\box0}}
{\setbox0=\hbox{$\scriptstyle\rm C$}\hbox{\hbox
to0pt{\kern0.4\wd0\vrule height0.9\ht0\hss}\box0}}
{\setbox0=\hbox{$\scriptscriptstyle\rm C$}\hbox{\hbox
to0pt{\kern0.4\wd0\vrule height0.9\ht0\hss}\box0}}}}
\def\bbbq{{\mathchoice {\setbox0=\hbox{$\displaystyle\rm
Q$}\hbox{\raise
0.15\ht0\hbox to0pt{\kern0.4\wd0\vrule height0.8\ht0\hss}\box0}}
{\setbox0=\hbox{$\textstyle\rm Q$}\hbox{\raise
0.15\ht0\hbox to0pt{\kern0.4\wd0\vrule height0.8\ht0\hss}\box0}}
{\setbox0=\hbox{$\scriptstyle\rm Q$}\hbox{\raise
0.15\ht0\hbox to0pt{\kern0.4\wd0\vrule height0.7\ht0\hss}\box0}}
{\setbox0=\hbox{$\scriptscriptstyle\rm Q$}\hbox{\raise
0.15\ht0\hbox to0pt{\kern0.4\wd0\vrule height0.7\ht0\hss}\box0}}}}
\def\bbbt{{\mathchoice {\setbox0=\hbox{$\displaystyle\rm
T$}\hbox{\hbox to0pt{\kern0.3\wd0\vrule height0.9\ht0\hss}\box0}}
{\setbox0=\hbox{$\textstyle\rm T$}\hbox{\hbox
to0pt{\kern0.3\wd0\vrule height0.9\ht0\hss}\box0}}
{\setbox0=\hbox{$\scriptstyle\rm T$}\hbox{\hbox
to0pt{\kern0.3\wd0\vrule height0.9\ht0\hss}\box0}}
{\setbox0=\hbox{$\scriptscriptstyle\rm T$}\hbox{\hbox
to0pt{\kern0.3\wd0\vrule height0.9\ht0\hss}\box0}}}}
\def\bbbs{{\mathchoice
{\setbox0=\hbox{$\displaystyle     \rm S$}\hbox{\raise0.5\ht0\hbox
to0pt{\kern0.35\wd0\vrule height0.45\ht0\hss}\hbox
to0pt{\kern0.55\wd0\vrule height0.5\ht0\hss}\box0}}
{\setbox0=\hbox{$\textstyle        \rm S$}\hbox{\raise0.5\ht0\hbox
to0pt{\kern0.35\wd0\vrule height0.45\ht0\hss}\hbox
to0pt{\kern0.55\wd0\vrule height0.5\ht0\hss}\box0}}
{\setbox0=\hbox{$\scriptstyle      \rm S$}\hbox{\raise0.5\ht0\hbox
to0pt{\kern0.35\wd0\vrule height0.45\ht0\hss}\raise0.05\ht0\hbox
to0pt{\kern0.5\wd0\vrule height0.45\ht0\hss}\box0}}
{\setbox0=\hbox{$\scriptscriptstyle\rm S$}\hbox{\raise0.5\ht0\hbox
to0pt{\kern0.4\wd0\vrule height0.45\ht0\hss}\raise0.05\ht0\hbox
to0pt{\kern0.55\wd0\vrule height0.45\ht0\hss}\box0}}}}
\def\bbbz{{\mathchoice {\hbox{$\sf\textstyle Z\kern-0.4em Z$}}
{\hbox{$\sf\textstyle Z\kern-0.4em Z$}}
{\hbox{$\sf\scriptstyle Z\kern-0.3em Z$}}
{\hbox{$\sf\scriptscriptstyle Z\kern-0.2em Z$}}}}


\ifprod@font
  \mathchardef\la="3\@xm2E
  \mathchardef\getsto="3\@xm1C
  \mathchardef\lid="3\@xm35
  \mathchardef\grole="3\@xm3F
  \mathchardef\loa="3\@xm2F
  \mathchardef\ga="3\@xm26
  \mathchardef\gid="3\@xm3D
  \mathchardef\leogr="3\@xm37
  \mathchardef\goa="3\@xm27
  \mathchardef\sq="0\@xm03
%
%
\def\diameter{{%
  \ifmmode
    \mathchoice
    {\ooalign{\hfil\hbox{$\displaystyle/$}\hfil\crcr
    {\lower.2ex\hbox{$\displaystyle\mathchar"20D$}}}}%
    {\ooalign{\hfil\hbox{$\textstyle/$}\hfil\crcr
    {\lower.2ex\hbox{$\textstyle\mathchar"20D$}}}}%
    {\ooalign{\hfil\hbox{$\scriptstyle/$}\hfil\crcr
    {\lower.1ex\hbox{$\scriptstyle\mathchar"20D$}}}}%
    {\ooalign{\hfil\hbox{$\scriptscriptstyle/$}\hfil\crcr
    {\lower.1ex\hbox{$\scriptscriptstyle\mathchar"20D$}}}}%
  \else
    {\ooalign{\hfil/\hfil\crcr\lower.2ex\hbox{\mathhexbox20D}}}%
  \fi
}}
%
%

\def\bbbc{{\Bbb{C}}}
\def\bbbq{{\Bbb{Q}}}
\def\bbbt{{\Bbb{T}}}
\def\bbbs{{\Bbb{S}}}
\def\bbbz{{\Bbb{Z}}}
\fi


\ifprod@font
\mathchardef\boxdot="2\@xm00
\mathchardef\boxplus="2\@xm01
\mathchardef\boxtimes="2\@xm02
\mathchardef\square="0\@xm03
\mathchardef\blacksquare="0\@xm04
\mathchardef\centerdot="2\@xm05
\mathchardef\lozenge="0\@xm06
\mathchardef\blacklozenge="0\@xm07
\mathchardef\circlearrowright="3\@xm08
\mathchardef\circlearrowleft="3\@xm09
\mathchardef\rightleftharpoons="3\@xm0A
\mathchardef\leftrightharpoons="3\@xm0B
\mathchardef\boxminus="2\@xm0C
\mathchardef\Vdash="3\@xm0D
\mathchardef\Vvdash="3\@xm0E
\mathchardef\vDash="3\@xm0F
\mathchardef\twoheadrightarrow="3\@xm10
\mathchardef\twoheadleftarrow="3\@xm11
\mathchardef\leftleftarrows="3\@xm12
\mathchardef\rightrightarrows="3\@xm13
\mathchardef\upuparrows="3\@xm14
\mathchardef\downdownarrows="3\@xm15
\mathchardef\upharpoonright="3\@xm16

\mathchardef\downharpoonright="3\@xm17
\mathchardef\upharpoonleft="3\@xm18
\mathchardef\downharpoonleft="3\@xm19
\mathchardef\rightarrowtail="3\@xm1A
\mathchardef\leftarrowtail="3\@xm1B
\mathchardef\leftrightarrows="3\@xm1C
\mathchardef\rightleftarrows="3\@xm1D
\mathchardef\Lsh="3\@xm1E
\mathchardef\Rsh="3\@xm1F
\mathchardef\rightsquigarrow="3\@xm20
\mathchardef\leftrightsquigarrow="3\@xm21
\mathchardef\looparrowleft="3\@xm22
\mathchardef\looparrowright="3\@xm23
\mathchardef\circeq="3\@xm24
\mathchardef\succsim="3\@xm25
\mathchardef\gtrsim="3\@xm26
\mathchardef\gtrapprox="3\@xm27
\mathchardef\multimap="3\@xm28
\mathchardef\therefore="3\@xm29
\mathchardef\because="3\@xm2A
\mathchardef\doteqdot="3\@xm2B

\mathchardef\triangleq="3\@xm2C
\mathchardef\precsim="3\@xm2D
\mathchardef\lesssim="3\@xm2E
\mathchardef\lessapprox="3\@xm2F
\mathchardef\eqslantless="3\@xm30
\mathchardef\eqslantgtr="3\@xm31
\mathchardef\curlyeqprec="3\@xm32
\mathchardef\curlyeqsucc="3\@xm33
\mathchardef\preccurlyeq="3\@xm34
\mathchardef\leqq="3\@xm35
\mathchardef\leqslant="3\@xm36
\mathchardef\lessgtr="3\@xm37
\mathchardef\backprime="0\@xm38
\mathchardef\risingdotseq="3\@xm3A
\mathchardef\fallingdotseq="3\@xm3B
\mathchardef\succcurlyeq="3\@xm3C
\mathchardef\geqq="3\@xm3D
\mathchardef\geqslant="3\@xm3E
\mathchardef\gtrless="3\@xm3F
\mathchardef\sqsubset="3\@xm40
\mathchardef\sqsupset="3\@xm41
\mathchardef\vartriangleright="3\@xm42
\mathchardef\vartriangleleft="3\@xm43
\mathchardef\trianglerighteq="3\@xm44
\mathchardef\trianglelefteq="3\@xm45
\mathchardef\bigstar="0\@xm46
\mathchardef\between="3\@xm47
\mathchardef\blacktriangledown="0\@xm48
\mathchardef\blacktriangleright="3\@xm49
\mathchardef\blacktriangleleft="3\@xm4A
\mathchardef\vartriangle="0\@xm4D
\mathchardef\blacktriangle="0\@xm4E
\mathchardef\triangledown="0\@xm4F
\mathchardef\eqcirc="3\@xm50
\mathchardef\lesseqgtr="3\@xm51
\mathchardef\gtreqless="3\@xm52
\mathchardef\lesseqqgtr="3\@xm53
\mathchardef\gtreqqless="3\@xm54
\mathchardef\Rrightarrow="3\@xm56
\mathchardef\Lleftarrow="3\@xm57
\mathchardef\veebar="2\@xm59
\mathchardef\barwedge="2\@xm5A
\mathchardef\doublebarwedge="2\@xm5B
\mathchardef\angle="0\@xm5C
\mathchardef\measuredangle="0\@xm5D
\mathchardef\sphericalangle="0\@xm5E
\mathchardef\varpropto="3\@xm5F
\mathchardef\smallsmile="3\@xm60
\mathchardef\smallfrown="3\@xm61
\mathchardef\Subset="3\@xm62
\mathchardef\Supset="3\@xm63
\mathchardef\Cup="2\@xm64

\mathchardef\Cap="2\@xm65

\mathchardef\curlywedge="2\@xm66
\mathchardef\curlyvee="2\@xm67
\mathchardef\leftthreetimes="2\@xm68
\mathchardef\rightthreetimes="2\@xm69
\mathchardef\subseteqq="3\@xm6A
\mathchardef\supseteqq="3\@xm6B
\mathchardef\bumpeq="3\@xm6C
\mathchardef\Bumpeq="3\@xm6D
\mathchardef\lll="3\@xm6E

\mathchardef\ggg="3\@xm6F

\mathchardef\circledS="0\@xm73
\mathchardef\pitchfork="3\@xm74
\mathchardef\dotplus="2\@xm75
\mathchardef\backsim="3\@xm76
\mathchardef\backsimeq="3\@xm77
\mathchardef\complement="0\@xm7B
\mathchardef\intercal="2\@xm7C
\mathchardef\circledcirc="2\@xm7D
\mathchardef\circledast="2\@xm7E
\mathchardef\circleddash="2\@xm7F
\def\ulcorner{\delimiter"4\@xm70\@xm70 }
\def\urcorner{\delimiter"5\@xm71\@xm71 }
\def\llcorner{\delimiter"4\@xm78\@xm78 }
\def\lrcorner{\delimiter"5\@xm79\@xm79 }
\def\yen{\mathhexbox\@xm55 }
\def\checkmark{\mathhexbox\@xm58 }
\def\circledR{\mathhexbox\@xm72 }
\def\maltese{\mathhexbox\@xm7A }
\mathchardef\lvertneqq="3\@ym00
\mathchardef\gvertneqq="3\@ym01
\mathchardef\nleq="3\@ym02
\mathchardef\ngeq="3\@ym03
\mathchardef\nless="3\@ym04
\mathchardef\ngtr="3\@ym05
\mathchardef\nprec="3\@ym06
\mathchardef\nsucc="3\@ym07
\mathchardef\lneqq="3\@ym08
\mathchardef\gneqq="3\@ym09
\mathchardef\nleqslant="3\@ym0A
\mathchardef\ngeqslant="3\@ym0B
\mathchardef\lneq="3\@ym0C
\mathchardef\gneq="3\@ym0D
\mathchardef\npreceq="3\@ym0E
\mathchardef\nsucceq="3\@ym0F
\mathchardef\precnsim="3\@ym10
\mathchardef\succnsim="3\@ym11
\mathchardef\lnsim="3\@ym12
\mathchardef\gnsim="3\@ym13
\mathchardef\nleqq="3\@ym14
\mathchardef\ngeqq="3\@ym15
\mathchardef\precneqq="3\@ym16
\mathchardef\succneqq="3\@ym17
\mathchardef\precnapprox="3\@ym18
\mathchardef\succnapprox="3\@ym19
\mathchardef\lnapprox="3\@ym1A
\mathchardef\gnapprox="3\@ym1B
\mathchardef\nsim="3\@ym1C
\mathchardef\ncong="3\@ym1D

\mathchardef\varsubsetneq="3\@ym20
\mathchardef\varsupsetneq="3\@ym21
\mathchardef\nsubseteqq="3\@ym22
\mathchardef\nsupseteqq="3\@ym23
\mathchardef\subsetneqq="3\@ym24
\mathchardef\supsetneqq="3\@ym25
\mathchardef\varsubsetneqq="3\@ym26
\mathchardef\varsupsetneqq="3\@ym27
\mathchardef\subsetneq="3\@ym28
\mathchardef\supsetneq="3\@ym29
\mathchardef\nsubseteq="3\@ym2A
\mathchardef\nsupseteq="3\@ym2B
\mathchardef\nparallel="3\@ym2C
\mathchardef\nmid="3\@ym2D
\mathchardef\nshortmid="3\@ym2E
\mathchardef\nshortparallel="3\@ym2F
\mathchardef\nvdash="3\@ym30
\mathchardef\nVdash="3\@ym31
\mathchardef\nvDash="3\@ym32
\mathchardef\nVDash="3\@ym33
\mathchardef\ntrianglerighteq="3\@ym34
\mathchardef\ntrianglelefteq="3\@ym35
\mathchardef\ntriangleleft="3\@ym36
\mathchardef\ntriangleright="3\@ym37
\mathchardef\nleftarrow="3\@ym38
\mathchardef\nrightarrow="3\@ym39
\mathchardef\nLeftarrow="3\@ym3A
\mathchardef\nRightarrow="3\@ym3B
\mathchardef\nLeftrightarrow="3\@ym3C
\mathchardef\nleftrightarrow="3\@ym3D
\mathchardef\divideontimes="2\@ym3E
\mathchardef\varnothing="0\@ym3F
\mathchardef\nexists="0\@ym40
\mathchardef\mho="0\@ym66
\mathchardef\eth="0\@ym67
\mathchardef\eqsim="3\@ym68
\mathchardef\beth="0\@ym69
\mathchardef\gimel="0\@ym6A
\mathchardef\daleth="0\@ym6B
\mathchardef\lessdot="3\@ym6C
\mathchardef\gtrdot="3\@ym6D
\mathchardef\ltimes="2\@ym6E
\mathchardef\rtimes="2\@ym6F
\mathchardef\shortmid="3\@ym70
\mathchardef\shortparallel="3\@ym71
\mathchardef\smallsetminus="2\@ym72
\mathchardef\thicksim="3\@ym73
\mathchardef\thickapprox="3\@ym74
\mathchardef\approxeq="3\@ym75
\mathchardef\succapprox="3\@ym76
\mathchardef\precapprox="3\@ym77
\mathchardef\curvearrowleft="3\@ym78
\mathchardef\curvearrowright="3\@ym79
\mathchardef\digamma="0\@ym7A
\mathchardef\varkappa="0\@ym7B
\mathchardef\hslash="0\@ym7D
\mathchardef\hbar="0\@ym7E
\mathchardef\backepsilon="3\@ym7F


\def\Bbb{\ifmmode\let\next\Bbb@\else
\def\next{\errmessage{Use \string\Bbb\space only in math mode}}\fi\next}
\def\Bbb@#1{{\Bbb@@{#1}}}
\def\Bbb@@#1{\fam\ymfam#1}
\fi


\def\Nulle{0} 
\def\Afe{1}   
\def\Hae{2}   
\def\Hbe{3}   
\def\Hce{4}   
\def\Hde{5}   


\newcount\LastMac       \LastMac=\Nulle

\newskip\half      \half=5.5pt plus 1.5pt minus 2.25pt
\newskip\one       \one=11pt plus 3pt minus 5.5pt
\newskip\onehalf   \onehalf=16.5pt plus 5.5pt minus 8.25pt
\newskip\two       \two=22pt plus 5.5pt minus 11pt

\def\Half{\addvspace{\half}}
\def\One{\addvspace{\one}}
\def\OneHalf{\addvspace{\onehalf}}
\def\Two{\addvspace{\two}}


\def\Raggedright{
  \rightskip=\z@ plus \hsize\relax
}

\def\Fullout{
  \rightskip=\z@\relax
}

\def\Hang#1#2{
  \hangindent=#1%
  \hangafter=#2\relax
}


\newif\ifsp@page
\def\pagestyle#1{\csname ps@#1\endcsname}
\def\thispagestyle#1{\global\sp@pagetrue\gdef\sp@type{#1}}

\def\ps@titlepage{%
  \def\@oddhead{\eightpoint\noindent \the\CatchLine
    \ifprod@font\else\qquad Printed\ \today\fi \hfil}%
  \let\@evenhead=\@oddhead
}

\def\ps@headings{%
  \def\@oddhead{\elevenpoint\it\noindent
    \hfill\the\RightHeader\hskip1.5em\rm\folio}%
  \def\@evenhead{\elevenpoint\noindent
    \folio\hskip1.5em\it\the\LeftHeader\hfill}%
}

\def\ps@plate{%
  \def\@oddhead{\eightpoint\noindent\plt@cap\hfil}%
  \def\@evenhead{\eightpoint\noindent\plt@cap\hfil}%
}



\def\title#1{
  \bgroup
    \vbox to 8pt{\vss}%
    \seventeenpoint
    \Raggedright
    \noindent \strut{\bf #1}\par
  \egroup
}

\def\author#1{
  \bgroup
    \ifnum\LastMac=\Afe \OneHalf\else \vskip 21pt\fi
    \fourteenpoint
    \Raggedright
    \noindent \strut #1\par
    \vskip 3pt%
  \egroup
}

\def\affiliation#1{
  \bgroup
    \vskip -4pt%
    \eightpoint
    \Raggedright
    \noindent \strut {\it #1}\par
  \egroup
  \LastMac=\Afe\relax
}

\def\acceptedline#1{
  \bgroup
    \Two
    \eightpoint
    \Raggedright
    \noindent \strut #1\par
  \egroup
}

\long\def\abstract#1{%
  \bgroup
    \vskip 20pt%
    \everypar{\Hang{11pc}{0}}%
    \noindent{\ninebf ABSTRACT}\par
    \tenpoint
    \Fullout
    \noindent #1\par
  \egroup
}

\long\def\keywords#1{
  \bgroup
    \Half
    \everypar{\Hang{11pc}{0}}%
    \tenpoint
    \Fullout
    \noindent\hbox{\bf Key words:}\ #1\par
  \egroup
}


\def\maketitle{%
  \EndOpening
  \ifsinglecol \else \MakePage\fi
}



\def\Autonumber{
  \global\AutoNumbertrue  
}

\newif\ifAutoNumber \AutoNumberfalse
\newcount\Sec        
\newcount\SecSec
\newcount\SecSecSec

\Sec=\z@

\def\:{\let\@sptoken= } \:  
\def\:{\@xifnch} \expandafter\def\: {\futurelet\@tempc\@ifnch}

\def\@ifnextchar#1#2#3{%
  \let\@tempMACe #1%
  \def\@tempMACa{#2}%
  \def\@tempMACb{#3}%
  \futurelet \@tempMACc\@ifnch%
}

\def\@ifnch{%
\ifx \@tempMACc \@sptoken%
  \let\@tempMACd\@xifnch%
\else%
  \ifx \@tempMACc \@tempMACe%
    \let\@tempMACd\@tempMACa%
  \else%
    \let\@tempMACd\@tempMACb%
  \fi%
\fi%
\@tempMACd%
}

\def\@ifstar#1#2{\@ifnextchar *{\def\@tempMACa*{#1}\@tempMACa}{#2}}

\newskip\@tempskipb

\def\addvspace#1{%
  \ifvmode\else \endgraf\fi%
  \ifdim\lastskip=\z@%
    \vskip #1\relax%
  \else%
    \@tempskipb#1\relax\@xaddvskip%
  \fi%
}

\def\@xaddvskip{%
  \ifdim\lastskip<\@tempskipb%
    \vskip-\lastskip%
    \vskip\@tempskipb\relax%
  \else%
    \ifdim\@tempskipb<\z@%
      \ifdim\lastskip<\z@ \else%
        \advance\@tempskipb\lastskip%
        \vskip-\lastskip\vskip\@tempskipb%
      \fi%
    \fi%
  \fi%
}

\newskip\@tmpSKIP

\def\addpen#1{%
  \ifvmode
    \if@nobreak
    \else
      \ifdim\lastskip=\z@
        \penalty#1\relax
      \else
        \@tmpSKIP=\lastskip
        \vskip -\lastskip
        \penalty#1\vskip\@tmpSKIP
      \fi
    \fi
  \fi
}

\newcount\@clubpen   \@clubpen=\clubpenalty
\newif\if@nobreak    \@nobreakfalse

\def\@noafterindent{%
  \global\@nobreaktrue
  \everypar{\if@nobreak
              \global\@nobreakfalse
              \clubpenalty \@M
              {\setbox\z@\lastbox}%
              \LastMac=\Nulle\relax%
            \else
              \clubpenalty \@clubpen
              \everypar{}%
            \fi}
}

\newcount\gds@cbrk   \gds@cbrk=-300

\def\@nohdbrk{\interlinepenalty \@M\relax}

\let\@par=\par
\def\@restorepar{\def\par{\@par}}

\newif\if@endpe   \@endpefalse
 
\def\@doendpe{\@endpetrue \@nobreakfalse \LastMac=\Nulle\relax%
     \def\par{\@restorepar\everypar{}\par\@endpefalse}%
              \everypar{\setbox\z@\lastbox\everypar{}\@endpefalse}%
}

\def\section{\@ifstar{\@ssection}{\@section}}

\def\@section#1{
  \if@nobreak
    \everypar{}%
    \ifnum\LastMac=\Hae \addvspace{\half}\fi
  \else
    \addpen{\gds@cbrk}%
    \addvspace{\two}%
  \fi
  \bgroup
    \ninepoint\bf
    \Raggedright
    \ifAutoNumber
      \global\advance\Sec \@ne
      \noindent\@nohdbrk\number\Sec\hskip 1pc \uppercase{#1}\par
      \global\SecSec=\z@
    \else
      \noindent\@nohdbrk\uppercase{#1}\par
    \fi
  \egroup
  \nobreak
  \vskip\half
  \nobreak
  \@noafterindent
  \LastMac=\Hae\relax
}

\def\@ssection#1{
  \if@nobreak
    \everypar{}%
    \ifnum\LastMac=\Hae \addvspace{\half}\fi
  \else
    \addpen{\gds@cbrk}%
    \addvspace{\two}%
  \fi
  \bgroup
    \ninepoint\bf
    \Raggedright
    \noindent\@nohdbrk\uppercase{#1}\par
  \egroup
  \nobreak
  \vskip\half
  \nobreak
  \@noafterindent
  \LastMac=\Hae\relax
}

\def\subsection#1{
  \if@nobreak
    \everypar{}%
    \ifnum\LastMac=\Hae \addvspace{1pt plus 1pt minus .5pt}\fi
  \else
    \addpen{\gds@cbrk}%
    \addvspace{\onehalf}%
  \fi
  \bgroup
    \ninepoint\bf
    \Raggedright
    \ifAutoNumber
      \global\advance\SecSec \@ne
      \noindent\@nohdbrk\number\Sec.\number\SecSec \hskip 1pc\relax #1\par
      \global\SecSecSec=\z@
    \else
      \noindent\@nohdbrk #1\par
    \fi
  \egroup
  \nobreak
  \vskip\half
  \nobreak
  \@noafterindent
  \LastMac=\Hbe\relax
}

\def\subsubsection#1{
  \if@nobreak
    \everypar{}%
    \ifnum\LastMac=\Hbe \addvspace{1pt plus 1pt minus .5pt}\fi
  \else
    \addpen{\gds@cbrk}%
    \addvspace{\onehalf}%
  \fi
  \bgroup
    \ninepoint\it
    \Raggedright
    \ifAutoNumber
      \global\advance\SecSecSec \@ne
      \noindent\@nohdbrk\number\Sec.\number\SecSec.\number\SecSecSec
        \hskip 1pc\relax #1\par
    \else
      \noindent\@nohdbrk #1\par
    \fi
  \egroup
  \nobreak
  \vskip\half
  \nobreak
  \@noafterindent
  \LastMac=\Hce\relax
}

\def\paragraph#1{
  \if@nobreak
    \everypar{}%
  \else
    \addpen{\gds@cbrk}%
    \addvspace{\one}%
  \fi%
  \bgroup%
    \ninepoint\it
    \noindent #1\ \nobreak%
  \egroup
  \LastMac=\Hde\relax
  \ignorespaces
}




\def\beginlist{%
  \par\if@nobreak \else\addvspace{\half}\fi%
  \bgroup%
    \ninepoint
    \let\item=\list@item%
}

\def\list@item{%
  \par\noindent\hskip 1em\relax%
  \ignorespaces%
}

\def\endlist{\par\egroup\addvspace{\half}\@doendpe}


\def\beginrefs{%
  \par
  \bgroup
    \eightpoint
    \Raggedright
    \let\bibitem=\bib@item
}

\def\bib@item{%
  \par\parindent=1.5em\Hang{1.5em}{1}%
  \everypar={\Hang{1.5em}{1}\ignorespaces}%
  \noindent\ignorespaces
}

\def\endrefs{\par\egroup\@doendpe}


\newtoks\CatchLine

\def\@journal{Mon.\ Not.\ R.\ Astron.\ Soc.\ }  
\def\@pubyear{1996}        
\def\@pagerange{000--000}  
\def\@volume{000}          
\def\@microfiche{}         %

\def\pubyear#1{\gdef\@pubyear{#1}\@makecatchline}
\def\pagerange#1{\gdef\@pagerange{#1}\@makecatchline}
\def\volume#1{\gdef\@volume{#1}\@makecatchline}
\def\microfiche#1{\gdef\@microfiche{and Microfiche\ #1}\@makecatchline}

\def\@makecatchline{%
  \global\CatchLine{%
    {\rm \@journal {\bf \@volume},\ \@pagerange\ (\@pubyear)\ \@microfiche}}%
}

\@makecatchline 

\newtoks\LeftHeader
\def\shortauthor#1{
  \global\LeftHeader{#1}%
}

\newtoks\RightHeader
\def\shorttitle#1{
  \global\RightHeader{#1}%
}

\def\PageHead{
  \begingroup
    \ifsp@page
      \csname ps@\sp@type\endcsname
      \global\sp@pagefalse
    \fi
    \ifodd\pageno
      \let\the@head=\@oddhead
    \else
      \let\the@head=\@evenhead
    \fi
    \vbox to \z@{\vskip-22.5\p@%
      \hbox to \PageWidth{\vbox to8.5\p@{}%
        \the@head
      }%
    \vss}%
  \endgroup
  \nointerlineskip
}

\def\today{%
  \number\day\space
  \ifcase\month\or January\or February\or March\or April\or May\or June\or
    July\or August\or September\or October\or November\or December\fi
  \space\number\year%
}

\def\PageFoot{} 

\def\authorcomment#1{%
  \gdef\PageFoot{%
    \nointerlineskip%
    \vbox to 22pt{\vfil%
      \hbox to \PageWidth{\elevenpoint\noindent \hfil #1 \hfil}}%
  }%
}


\newif\ifplate@page
\newbox\plt@box

\def\beginplatepage{%
  \let\plate=\plate@head
  \let\caption=\fig@caption
  \global\setbox\plt@box=\vbox\bgroup
  \TEMPDIMEN=\PageWidth 
  \hsize=\PageWidth\relax
}

\def\endplatepage{\par\egroup\global\plate@pagetrue}
\def\plate@head#1{\gdef\plt@cap{#1}}


\def\letters{%
  \gdef\folio{\ifnum\pageno<\z@ L\romannumeral-\pageno
    \else L\number\pageno \fi}%
}


\everydisplay{\displaysetup}

\newif\ifeqno
\newif\ifleqno

\def\displaysetup#1$${%
 \displaytest#1\eqno\eqno\displaytest
}

\def\displaytest#1\eqno#2\eqno#3\displaytest{%
 \if!#3!\ldisplaytest#1\leqno\leqno\ldisplaytest
 \else\eqnotrue\leqnofalse\def\eqn{#2}\def\eq{#1}\fi
 \generaldisplay$$}

\def\ldisplaytest#1\leqno#2\leqno#3\ldisplaytest{%
 \def\eq{#1}%
 \if!#3!\eqnofalse\else\eqnotrue\leqnotrue
  \def\eqn{#2}\fi}

\def\generaldisplay{%
\ifeqno \ifleqno 
   \hbox to \hsize{\noindent
     $\displaystyle\eq$\hfil$\displaystyle\eqn$}
  \else
    \hbox to \hsize{\noindent
     $\displaystyle\eq$\hfil$\displaystyle\eqn$}
  \fi
 \else
 \hbox to \hsize{\vbox{\noindent
  $\displaystyle\eq$\hfil}}
 \fi
}


\def\@notice{%
  \par\Two%
  \noindent{\b@ls{11pt}\ninerm This paper has been produced using the
    Blackwell Scientific Publications \TeX\ macros.\par}%
}

\outer\def\bye{\@notice\par\vfill\supereject\end}


\def\start@mess{%
  Monthly notices of the RAS journal style (\@typeface)\space
    v\@version,\space \@verdate.%
}

\everyjob{\Warn{\start@mess}}



\newif\if@debug \@debugfalse  

\def\Print#1{\if@debug\immediate\write16{#1}\else \fi}
\def\Warn#1{\immediate\write16{#1}}
\def\wlog#1{}

\newcount\Iteration 

\def\Single{0} \def\Double{1}                 
\def\Figure{0} \def\Table{1}                  

\def\InStack{0}  
\def\InZoneA{1}
\def\InZoneB{2}
\def\InZoneC{3}

\newcount\TEMPCOUNT 
\newdimen\TEMPDIMEN 
\newbox\TEMPBOX     
\newbox\VOIDBOX     

\newcount\LengthOfStack 
\newcount\MaxItems      
\newcount\StackPointer
\newcount\Point         
\newcount\NextFigure    
\newcount\NextTable     
\newcount\NextItem      

\newcount\StatusStack   
\newcount\NumStack      
\newcount\TypeStack     
\newcount\SpanStack     
\newcount\BoxStack      

\newcount\ItemSTATUS    
\newcount\ItemNUMBER    
\newcount\ItemTYPE      
\newcount\ItemSPAN      
\newbox\ItemBOX         
\newdimen\ItemSIZE      

\newdimen\PageHeight    
\newdimen\TextLeading   
\newdimen\Feathering    
\newcount\LinesPerPage  
\newdimen\ColumnWidth   
\newdimen\ColumnGap     
\newdimen\PageWidth     
\newdimen\BodgeHeight   
\newcount\Leading       

\newdimen\ZoneBSize  
\newdimen\TextSize   
\newbox\ZoneABOX     
\newbox\ZoneBBOX     
\newbox\ZoneCBOX     

\newif\ifFirstSingleItem
\newif\ifFirstZoneA
\newif\ifMakePageInComplete
\newif\ifMoreFigures \MoreFiguresfalse 
\newif\ifMoreTables  \MoreTablesfalse  

\newif\ifFigInZoneB 
\newif\ifFigInZoneC 
\newif\ifTabInZoneB 
\newif\ifTabInZoneC

\newif\ifZoneAFullPage

\newbox\MidBOX    
\newbox\LeftBOX
\newbox\RightBOX
\newbox\PageBOX   

\newif\ifLeftCOL  
\LeftCOLtrue

\newdimen\ZoneBAdjust

\newcount\ItemFits
\def\Yes{1}
\def\No{2}


\MaxItems=15
\NextFigure=\z@        
\NextTable=\@ne

\BodgeHeight=6pt
\TextLeading=11pt    
\Leading=11
\Feathering=\z@      
\LinesPerPage=61     
\topskip=\TextLeading
\ColumnWidth=20pc    
\ColumnGap=2pc       

\newskip\ItemSepamount  
\ItemSepamount=\TextLeading plus \TextLeading minus 4pt

\parskip=\z@ plus .1pt
\parindent=18pt
\widowpenalty=\z@
\clubpenalty=10000
\tolerance=1500
\hbadness=1500
\abovedisplayskip=6pt plus 2pt minus 2pt
\belowdisplayskip=6pt plus 2pt minus 2pt
\abovedisplayshortskip=6pt plus 2pt minus 2pt
\belowdisplayshortskip=6pt plus 2pt minus 2pt

\ninepoint 


\PageHeight=682pt

\PageWidth=2\ColumnWidth
\advance\PageWidth by \ColumnGap

\pagestyle{headings}




\newcount\DUMMY \StatusStack=\allocationnumber
\newcount\DUMMY \newcount\DUMMY \newcount\DUMMY 
\newcount\DUMMY \newcount\DUMMY \newcount\DUMMY 
\newcount\DUMMY \newcount\DUMMY \newcount\DUMMY
\newcount\DUMMY \newcount\DUMMY \newcount\DUMMY 
\newcount\DUMMY \newcount\DUMMY \newcount\DUMMY

\newcount\DUMMY \NumStack=\allocationnumber
\newcount\DUMMY \newcount\DUMMY \newcount\DUMMY 
\newcount\DUMMY \newcount\DUMMY \newcount\DUMMY 
\newcount\DUMMY \newcount\DUMMY \newcount\DUMMY 
\newcount\DUMMY \newcount\DUMMY \newcount\DUMMY 
\newcount\DUMMY \newcount\DUMMY \newcount\DUMMY

\newcount\DUMMY \TypeStack=\allocationnumber
\newcount\DUMMY \newcount\DUMMY \newcount\DUMMY 
\newcount\DUMMY \newcount\DUMMY \newcount\DUMMY 
\newcount\DUMMY \newcount\DUMMY \newcount\DUMMY 
\newcount\DUMMY \newcount\DUMMY \newcount\DUMMY 
\newcount\DUMMY \newcount\DUMMY \newcount\DUMMY

\newcount\DUMMY \SpanStack=\allocationnumber
\newcount\DUMMY \newcount\DUMMY \newcount\DUMMY 
\newcount\DUMMY \newcount\DUMMY \newcount\DUMMY 
\newcount\DUMMY \newcount\DUMMY \newcount\DUMMY 
\newcount\DUMMY \newcount\DUMMY \newcount\DUMMY 
\newcount\DUMMY \newcount\DUMMY \newcount\DUMMY

\newbox\DUMMY   \BoxStack=\allocationnumber
\newbox\DUMMY   \newbox\DUMMY \newbox\DUMMY 
\newbox\DUMMY   \newbox\DUMMY \newbox\DUMMY 
\newbox\DUMMY   \newbox\DUMMY \newbox\DUMMY 
\newbox\DUMMY   \newbox\DUMMY \newbox\DUMMY 
\newbox\DUMMY   \newbox\DUMMY \newbox\DUMMY

\def\wlog{\immediate\write\m@ne}


\def\GetItemAll#1{%
 \GetItemSTATUS{#1}
 \GetItemNUMBER{#1}
 \GetItemTYPE{#1}
 \GetItemSPAN{#1}
 \GetItemBOX{#1}
}

\def\GetItemSTATUS#1{%
 \Point=\StatusStack
 \advance\Point by #1
 \global\ItemSTATUS=\count\Point
}

\def\GetItemNUMBER#1{%
 \Point=\NumStack
 \advance\Point by #1
 \global\ItemNUMBER=\count\Point
}

\def\GetItemTYPE#1{%
 \Point=\TypeStack
 \advance\Point by #1
 \global\ItemTYPE=\count\Point
}

\def\GetItemSPAN#1{%
 \Point\SpanStack
 \advance\Point by #1
 \global\ItemSPAN=\count\Point
}

\def\GetItemBOX#1{%
 \Point=\BoxStack
 \advance\Point by #1
 \global\setbox\ItemBOX=\vbox{\copy\Point}
 \global\ItemSIZE=\ht\ItemBOX
 \global\advance\ItemSIZE by \dp\ItemBOX
 \TEMPCOUNT=\ItemSIZE
 \divide\TEMPCOUNT by \Leading
 \divide\TEMPCOUNT by 65536
 \advance\TEMPCOUNT \@ne
 \ItemSIZE=\TEMPCOUNT pt
 \global\multiply\ItemSIZE by \Leading
}


\def\JoinStack{%
 \ifnum\LengthOfStack=\MaxItems 
  \Warn{WARNING: Stack is full...some items will be lost!}
 \else
  \Point=\StatusStack
  \advance\Point by \LengthOfStack
  \global\count\Point=\ItemSTATUS
  \Point=\NumStack
  \advance\Point by \LengthOfStack
  \global\count\Point=\ItemNUMBER
  \Point=\TypeStack
  \advance\Point by \LengthOfStack
  \global\count\Point=\ItemTYPE
  \Point\SpanStack
  \advance\Point by \LengthOfStack
  \global\count\Point=\ItemSPAN
  \Point=\BoxStack
  \advance\Point by \LengthOfStack
  \global\setbox\Point=\vbox{\copy\ItemBOX}
  \global\advance\LengthOfStack \@ne
  \ifnum\ItemTYPE=\Figure 
   \global\MoreFigurestrue
  \else
   \global\MoreTablestrue
  \fi
 \fi
}


\def\LeaveStack#1{%
 {\Iteration=#1
 \loop
 \ifnum\Iteration<\LengthOfStack
  \advance\Iteration \@ne
  \GetItemSTATUS{\Iteration}
   \advance\Point by \m@ne
   \global\count\Point=\ItemSTATUS
  \GetItemNUMBER{\Iteration}
   \advance\Point by \m@ne
   \global\count\Point=\ItemNUMBER
  \GetItemTYPE{\Iteration}
   \advance\Point by \m@ne
   \global\count\Point=\ItemTYPE
  \GetItemSPAN{\Iteration}
   \advance\Point by \m@ne
   \global\count\Point=\ItemSPAN
  \GetItemBOX{\Iteration}
   \advance\Point by \m@ne
   \global\setbox\Point=\vbox{\copy\ItemBOX}
 \repeat}
 \global\advance\LengthOfStack by \m@ne
}


\newif\ifStackNotClean

\def\CleanStack{%
 \StackNotCleantrue
 {\Iteration=\z@
  \loop
   \ifStackNotClean
    \GetItemSTATUS{\Iteration}
    \ifnum\ItemSTATUS=\InStack
     \advance\Iteration \@ne
     \else
      \LeaveStack{\Iteration}
    \fi
   \ifnum\LengthOfStack<\Iteration
    \StackNotCleanfalse
   \fi
 \repeat}
}


\def\FindItem#1#2{%
 \global\StackPointer=\m@ne 
 {\Iteration=\z@
  \loop
  \ifnum\Iteration<\LengthOfStack
   \GetItemSTATUS{\Iteration}
   \ifnum\ItemSTATUS=\InStack
    \GetItemTYPE{\Iteration}
    \ifnum\ItemTYPE=#1
     \GetItemNUMBER{\Iteration}
     \ifnum\ItemNUMBER=#2
      \global\StackPointer=\Iteration
      \Iteration=\LengthOfStack 
     \fi
    \fi
   \fi
  \advance\Iteration \@ne
 \repeat}
}


\def\FindNext{%
 \global\StackPointer=\m@ne 
 {\Iteration=\z@
  \loop
  \ifnum\Iteration<\LengthOfStack
   \GetItemSTATUS{\Iteration}
   \ifnum\ItemSTATUS=\InStack
    \GetItemTYPE{\Iteration}
   \ifnum\ItemTYPE=\Figure
    \ifMoreFigures
      \global\NextItem=\Figure
      \global\StackPointer=\Iteration
      \Iteration=\LengthOfStack 
    \fi
   \fi
   \ifnum\ItemTYPE=\Table
    \ifMoreTables
      \global\NextItem=\Table
      \global\StackPointer=\Iteration
      \Iteration=\LengthOfStack 
    \fi
   \fi
  \fi
  \advance\Iteration \@ne
 \repeat}
}


\def\ChangeStatus#1#2{%
 \Point=\StatusStack
 \advance\Point by #1
 \global\count\Point=#2
}



\def\Zone{\InZoneA}

\ZoneBAdjust=\z@

\def\MakePage{
 \global\ZoneBSize=\PageHeight
 \global\TextSize=\ZoneBSize
 \global\ZoneAFullPagefalse
 \global\topskip=\TextLeading
 \MakePageInCompletetrue
 \MoreFigurestrue
 \MoreTablestrue
 \FigInZoneBfalse
 \FigInZoneCfalse
 \TabInZoneBfalse
 \TabInZoneCfalse
 \global\FirstSingleItemtrue
 \global\FirstZoneAtrue
 \global\setbox\ZoneABOX=\box\VOIDBOX
 \global\setbox\ZoneBBOX=\box\VOIDBOX
 \global\setbox\ZoneCBOX=\box\VOIDBOX
 \loop
  \ifMakePageInComplete
 \FindNext
 \ifnum\StackPointer=\m@ne
  \NextItem=\m@ne
  \MoreFiguresfalse
  \MoreTablesfalse
 \fi
 \ifnum\NextItem=\Figure
   \FindItem{\Figure}{\NextFigure}
   \ifnum\StackPointer=\m@ne \global\MoreFiguresfalse
   \else
    \GetItemSPAN{\StackPointer}
    \ifnum\ItemSPAN=\Single \def\Zone{\InZoneB}\relax
     \ifFigInZoneC \global\MoreFiguresfalse\fi
    \else
     \def\Zone{\InZoneA}
     \ifFigInZoneB \def\Zone{\InZoneC}\fi
    \fi
   \fi
   \ifMoreFigures\Print{}\FigureItems\fi
 \fi
\ifnum\NextItem=\Table
   \FindItem{\Table}{\NextTable}
   \ifnum\StackPointer=\m@ne \global\MoreTablesfalse
   \else
    \GetItemSPAN{\StackPointer}
    \ifnum\ItemSPAN=\Single\relax
     \ifTabInZoneC \global\MoreTablesfalse\fi
    \else
     \def\Zone{\InZoneA}
     \ifTabInZoneB \def\Zone{\InZoneC}\fi
    \fi
   \fi
   \ifMoreTables\Print{}\TableItems\fi
 \fi
   \MakePageInCompletefalse 
   \ifMoreFigures\MakePageInCompletetrue\fi
   \ifMoreTables\MakePageInCompletetrue\fi
 \repeat
 \ifZoneAFullPage
  \global\TextSize=\z@
  \global\ZoneBSize=\z@
  \global\vsize=\z@\relax
  \global\topskip=\z@\relax
  \vbox to \z@{\vss}
  \eject
 \else
 \global\advance\ZoneBSize by -\ZoneBAdjust
 \global\vsize=\ZoneBSize
 \global\hsize=\ColumnWidth
 \global\ZoneBAdjust=\z@
 \ifdim\TextSize<23pt
 \Warn{}
 \Warn{* Making column fall short: TextSize=\the\TextSize *}
 \vskip-\lastskip\eject\fi
 \fi
}

\def\MakeRightCol{
 \global\TextSize=\ZoneBSize
 \MakePageInCompletetrue
 \MoreFigurestrue
 \MoreTablestrue
 \global\FirstSingleItemtrue
 \global\setbox\ZoneBBOX=\box\VOIDBOX
 \def\Zone{\InZoneB}
 \loop
  \ifMakePageInComplete
 \FindNext
 \ifnum\StackPointer=\m@ne
  \NextItem=\m@ne
  \MoreFiguresfalse
  \MoreTablesfalse
 \fi
 \ifnum\NextItem=\Figure
   \FindItem{\Figure}{\NextFigure}
   \ifnum\StackPointer=\m@ne \MoreFiguresfalse
   \else
    \GetItemSPAN{\StackPointer}
    \ifnum\ItemSPAN=\Double\relax
     \MoreFiguresfalse\fi
   \fi
   \ifMoreFigures\Print{}\FigureItems\fi
 \fi
 \ifnum\NextItem=\Table
   \FindItem{\Table}{\NextTable}
   \ifnum\StackPointer=\m@ne \MoreTablesfalse
   \else
    \GetItemSPAN{\StackPointer}
    \ifnum\ItemSPAN=\Double\relax
     \MoreTablesfalse\fi
   \fi
   \ifMoreTables\Print{}\TableItems\fi
 \fi
   \MakePageInCompletefalse 
   \ifMoreFigures\MakePageInCompletetrue\fi
   \ifMoreTables\MakePageInCompletetrue\fi
 \repeat
 \ifZoneAFullPage
  \global\TextSize=\z@
  \global\ZoneBSize=\z@
  \global\vsize=\z@\relax
  \global\topskip=\z@\relax
  \vbox to \z@{\vss}
  \eject
 \else
 \global\vsize=\ZoneBSize
 \global\hsize=\ColumnWidth
 \ifdim\TextSize<23pt
 \Warn{}
 \Warn{* Making column fall short: TextSize=\the\TextSize *}
 \vskip-\lastskip\eject\fi
\fi
}

\def\FigureItems{
 \Print{Considering...}
 \ShowItem{\StackPointer}
 \GetItemBOX{\StackPointer} 
 \GetItemSPAN{\StackPointer}
  \CheckFitInZone 
  \ifnum\ItemFits=\Yes
   \ifnum\ItemSPAN=\Single
     \ChangeStatus{\StackPointer}{\InZoneB} 
     \global\FigInZoneBtrue
     \ifFirstSingleItem
      \hbox{}\vskip-\BodgeHeight
     \global\advance\ItemSIZE by \TextLeading
     \fi
     \unvbox\ItemBOX\ItemSep
     \global\FirstSingleItemfalse
     \global\advance\TextSize by -\ItemSIZE
     \global\advance\TextSize by -\TextLeading
   \else
    \ifFirstZoneA
     \global\advance\ItemSIZE by \TextLeading
     \global\FirstZoneAfalse\fi
    \global\advance\TextSize by -\ItemSIZE
    \global\advance\TextSize by -\TextLeading
    \global\advance\ZoneBSize by -\ItemSIZE
    \global\advance\ZoneBSize by -\TextLeading
    \ifFigInZoneB\relax
     \else
     \ifdim\TextSize<3\TextLeading
     \global\ZoneAFullPagetrue
     \fi
    \fi
    \ChangeStatus{\StackPointer}{\Zone}
    \ifnum\Zone=\InZoneC \global\FigInZoneCtrue\fi
  \fi
   \Print{TextSize=\the\TextSize}
   \Print{ZoneBSize=\the\ZoneBSize}
  \global\advance\NextFigure \@ne
   \Print{This figure has been placed.}
  \else
   \Print{No space available for this figure...holding over.}
   \Print{}
   \global\MoreFiguresfalse
  \fi
}

\def\TableItems{
 \Print{Considering...}
 \ShowItem{\StackPointer}
 \GetItemBOX{\StackPointer} 
 \GetItemSPAN{\StackPointer}
  \CheckFitInZone 
  \ifnum\ItemFits=\Yes
   \ifnum\ItemSPAN=\Single
    \ChangeStatus{\StackPointer}{\InZoneB}
     \global\TabInZoneBtrue
     \ifFirstSingleItem
      \hbox{}\vskip-\BodgeHeight
     \global\advance\ItemSIZE by \TextLeading
     \fi
     \unvbox\ItemBOX\ItemSep
     \global\FirstSingleItemfalse
     \global\advance\TextSize by -\ItemSIZE
     \global\advance\TextSize by -\TextLeading
   \else
    \ifFirstZoneA
    \global\advance\ItemSIZE by \TextLeading
    \global\FirstZoneAfalse\fi
    \global\advance\TextSize by -\ItemSIZE
    \global\advance\TextSize by -\TextLeading
    \global\advance\ZoneBSize by -\ItemSIZE
    \global\advance\ZoneBSize by -\TextLeading
    \ifFigInZoneB\relax
     \else
     \ifdim\TextSize<3\TextLeading
     \global\ZoneAFullPagetrue
     \fi
    \fi
    \ChangeStatus{\StackPointer}{\Zone}
    \ifnum\Zone=\InZoneC \global\TabInZoneCtrue\fi
   \fi
  \global\advance\NextTable \@ne
   \Print{This table has been placed.}
  \else
  \Print{No space available for this table...holding over.}
   \Print{}
   \global\MoreTablesfalse
  \fi
}


\def\CheckFitInZone{%
{\advance\TextSize by -\ItemSIZE
 \advance\TextSize by -\TextLeading
 \ifFirstSingleItem
  \advance\TextSize by \TextLeading
 \fi
 \ifnum\Zone=\InZoneA\relax
  \else \advance\TextSize by -\ZoneBAdjust
 \fi
 \ifdim\TextSize<3\TextLeading \global\ItemFits=\No
 \else \global\ItemFits=\Yes\fi}
}

\def\BeginOpening{%
  \thispagestyle{titlepage}%
  \global\setbox\ItemBOX=\vbox\bgroup%
    \hsize=\PageWidth%
    \hrule height \z@
    \ifsinglecol\vskip 6pt\fi 
}

\let\begintopmatter=\BeginOpening  

\def\EndOpening{%
  \One
  \egroup
  \ifsinglecol
    \box\ItemBOX%
    \vskip\TextLeading plus 2\TextLeading
    \@noafterindent
  \else
    \ItemNUMBER=\z@%
    \ItemTYPE=\Figure
    \ItemSPAN=\Double
    \ItemSTATUS=\InStack
    \JoinStack
  \fi
}


\newif\if@here  \@herefalse

\def\no@float{\global\@heretrue}
\let\nofloat=\relax 

\def\beginfigure{%
  \@ifstar{\global\@dfloattrue \@bfigure}{\global\@dfloatfalse \@bfigure}%
}

\def\@bfigure#1{%
  \par
  \if@dfloat
    \ItemSPAN=\Double
    \TEMPDIMEN=\PageWidth
  \else
    \ItemSPAN=\Single
    \TEMPDIMEN=\ColumnWidth
  \fi
  \ifsinglecol
    \TEMPDIMEN=\PageWidth
  \else
    \ItemSTATUS=\InStack
    \ItemNUMBER=#1%
    \ItemTYPE=\Figure
  \fi
  \bgroup
    \hsize=\TEMPDIMEN
    \global\setbox\ItemBOX=\vbox\bgroup
      \eightpoint\nostb@ls{10pt}%
      \let\caption=\fig@caption
      \ifsinglecol \let\nofloat=\no@float\fi
}

\def\fig@caption#1{%
  \vskip 5.5pt plus 6pt%
  \bgroup 
    \eightpoint\nostb@ls{10pt}%
    \setbox\TEMPBOX=\hbox{#1}%
    \ifdim\wd\TEMPBOX>\TEMPDIMEN
      \noindent \unhbox\TEMPBOX\par
    \else
      \hbox to \hsize{\hfil\unhbox\TEMPBOX\hfil}%
    \fi
  \egroup
}

\def\endfigure{%
  \par\egroup 
  \egroup
  \ifsinglecol
    \if@here \midinsert\global\@herefalse\else \topinsert\fi
      \unvbox\ItemBOX
    \endinsert
  \else
    \JoinStack
    \Print{Processing source for figure \the\ItemNUMBER}%
  \fi
}


\newbox\tab@cap@box
\def\tab@caption#1{\global\setbox\tab@cap@box=\hbox{#1\par}}

\newtoks\tab@txt@toks
\long\def\tab@txt#1{\global\tab@txt@toks={#1}\global\table@txttrue}

\newif\iftable@txt  \table@txtfalse
\newif\if@dfloat    \@dfloatfalse

\def\begintable{%
  \@ifstar{\global\@dfloattrue \@btable}{\global\@dfloatfalse \@btable}%
}

\def\@btable#1{%
  \par
  \if@dfloat
    \ItemSPAN=\Double
    \TEMPDIMEN=\PageWidth
  \else
    \ItemSPAN=\Single
    \TEMPDIMEN=\ColumnWidth
  \fi
  \ifsinglecol
    \TEMPDIMEN=\PageWidth
  \else
    \ItemSTATUS=\InStack
    \ItemNUMBER=#1%
    \ItemTYPE=\Table
  \fi
  \bgroup
    \eightpoint\nostb@ls{10pt}%
    \global\setbox\ItemBOX=\vbox\bgroup
      \let\caption=\tab@caption
      \let\tabletext=\tab@txt
      \ifsinglecol \let\nofloat=\no@float\fi
}

\def\endtable{%
  \par\egroup 
  \egroup
  \setbox\TEMPBOX=\hbox to \TEMPDIMEN{%
    \hss
    \vbox{%
      \hsize=\wd\ItemBOX
      \ifvoid\tab@cap@box
      \else
        \noindent\unhbox\tab@cap@box
        \vskip 5.5pt plus 6pt%
      \fi
      \box\ItemBOX
      \iftable@txt
        \vskip 10pt%
        \eightpoint\nostb@ls{10pt}%
        \noindent\the\tab@txt@toks
        \global\table@txtfalse
      \fi
    }%
    \hss
  }%
  \ifsinglecol
    \if@here \midinsert\global\@herefalse\else \topinsert\fi
      \box\TEMPBOX
    \endinsert
  \else
    \global\setbox\ItemBOX=\box\TEMPBOX
    \JoinStack
    \Print{Processing source for table \the\ItemNUMBER}%
  \fi
}

\def\UnloadZoneA{%
\FirstZoneAtrue
 \Iteration=\z@
  \loop
   \ifnum\Iteration<\LengthOfStack
    \GetItemSTATUS{\Iteration}
    \ifnum\ItemSTATUS=\InZoneA
     \GetItemBOX{\Iteration}
     \ifFirstZoneA \vbox to \BodgeHeight{\vfil}%
     \FirstZoneAfalse\fi
     \unvbox\ItemBOX\ItemSep
     \LeaveStack{\Iteration}
     \else
     \advance\Iteration \@ne
   \fi
 \repeat
}

\def\UnloadZoneC{%
\Iteration=\z@
  \loop
   \ifnum\Iteration<\LengthOfStack
    \GetItemSTATUS{\Iteration}
    \ifnum\ItemSTATUS=\InZoneC
     \GetItemBOX{\Iteration}
     \ItemSep\unvbox\ItemBOX
     \LeaveStack{\Iteration}
     \else
     \advance\Iteration \@ne
   \fi
 \repeat
}


\def\ShowItem#1{
  {\GetItemAll{#1}
  \Print{\the#1:
  {TYPE=\ifnum\ItemTYPE=\Figure Figure\else Table\fi}
  {NUMBER=\the\ItemNUMBER}
  {SPAN=\ifnum\ItemSPAN=\Single Single\else Double\fi}
  {SIZE=\the\ItemSIZE}}}
}

\def\ShowStack{%
 \Print{}
 \Print{LengthOfStack = \the\LengthOfStack}
 \ifnum\LengthOfStack=\z@ \Print{Stack is empty}\fi
 \Iteration=\z@
 \loop
 \ifnum\Iteration<\LengthOfStack
  \ShowItem{\Iteration}
  \advance\Iteration \@ne
 \repeat
}

\def\B#1#2{%
\hbox{\vrule\kern-0.4pt\vbox to #2{%
\hrule width #1\vfill\hrule}\kern-0.4pt\vrule}
}


\newif\ifsinglecol   \singlecolfalse

\def\onecolumn{%
  \global\output={\singlecoloutput}%
  \global\hsize=\PageWidth
  \global\vsize=\PageHeight
  \global\ColumnWidth=\hsize
  \global\TextLeading=12pt
  \global\Leading=12
  \global\singlecoltrue
  \global\let\onecolumn=\relax
  \global\let\footnote=\sing@footnote
  \global\let\vfootnote=\sing@vfootnote
  \ninepoint 
  \message{(Single column)}%
}

\def\singlecoloutput{%
  \shipout\vbox{\PageHead\pagebody\PageFoot}%
  \advancepageno
  \ifplate@page
    \shipout\vbox{%
      \sp@pagetrue
      \def\sp@type{plate}%
      \global\plate@pagefalse
      \PageHead\vbox to \PageHeight{\unvbox\plt@box\vfil}\PageFoot%
    }%
    \message{[plate]}%
    \advancepageno
  \fi
  \ifnum\outputpenalty>-\@MM \else\dosupereject\fi%
}

\def\ItemSep{\vskip\ItemSepamount\relax}

\def\ItemSepbreak{\par\ifdim\lastskip<\ItemSepamount
  \removelastskip\penalty-200\ItemSep\fi%
}


\let\@@endinsert=\endinsert 

\def\endinsert{\egroup 
  \if@mid \dimen@\ht\z@ \advance\dimen@\dp\z@ \advance\dimen@12\p@
    \advance\dimen@\pagetotal \advance\dimen@-\pageshrink
    \ifdim\dimen@>\pagegoal\@midfalse\p@gefalse\fi\fi
  \if@mid \ItemSep\box\z@\ItemSepbreak
  \else\insert\topins{\penalty100 
    \splittopskip\z@skip
    \splitmaxdepth\maxdimen \floatingpenalty\z@
    \ifp@ge \dimen@\dp\z@
    \vbox to\vsize{\unvbox\z@\kern-\dimen@}
    \else \box\z@\nobreak\ItemSep\fi}\fi\endgroup%
}


\def\gobbleone#1{}
\def\gobbletwo#1#2{}
\let\footnote=\gobbletwo 
\let\vfootnote=\gobbleone

\def\sing@footnote#1{\let\@sf\empty 
  \ifhmode\edef\@sf{\spacefactor\the\spacefactor}\/\fi
  \hbox{$^{\hbox{\eightpoint #1}}$}\@sf\sing@vfootnote{#1}%
}

\def\sing@vfootnote#1{\insert\footins\bgroup\eightpoint\b@ls{9pt}%
  \interlinepenalty\interfootnotelinepenalty
  \splittopskip\ht\strutbox 
  \splitmaxdepth\dp\strutbox \floatingpenalty\@MM
  \leftskip\z@skip \rightskip\z@skip \spaceskip\z@skip \xspaceskip\z@skip
  \noindent $^{\scriptstyle\hbox{#1}}$\hskip 4pt%
    \footstrut\futurelet\next\fo@t%
}

\def\footnoterule{\kern-3\p@ \hrule height \z@ \kern 3\p@}

\skip\footins=19.5pt plus 12pt minus 1pt
\count\footins=1000
\dimen\footins=\maxdimen


\def\landscape{%
  \global\TEMPDIMEN=\PageWidth
  \global\PageWidth=\PageHeight
  \global\PageHeight=\TEMPDIMEN
  \global\let\landscape=\relax
  \onecolumn
  \message{(landscape)}%
  \raggedbottom
}


\output{%
  \ifLeftCOL
    \global\setbox\LeftBOX=\vbox to \ZoneBSize{\box255\unvbox\ZoneBBOX}%
    \global\LeftCOLfalse
    \MakeRightCol
  \else
    \setbox\RightBOX=\vbox to \ZoneBSize{\box255\unvbox\ZoneBBOX}%
    \setbox\MidBOX=\hbox{\box\LeftBOX\hskip\ColumnGap\box\RightBOX}%
    \setbox\PageBOX=\vbox to \PageHeight{%
      \UnloadZoneA\box\MidBOX\UnloadZoneC}%
    \shipout\vbox{\PageHead\box\PageBOX\PageFoot}%
    \advancepageno
    \ifplate@page
      \shipout\vbox{%
        \sp@pagetrue
        \def\sp@type{plate}%
        \global\plate@pagefalse
        \PageHead\vbox to \PageHeight{\unvbox\plt@box\vfil}\PageFoot%
      }%
      \message{[plate]}%
      \advancepageno
    \fi
    \global\LeftCOLtrue
    \CleanStack
    \MakePage
  \fi
}


\Warn{\start@mess}


\catcode `\@=12 



%% file: mnextra.tex
\hoffset=-.5cm
\voffset=.5cm

\overfullrule=0pt

\def\chaphead{}
\def\today{\ifcase\month\or
 January\or February\or March\or April\or May\or June\or
 July\or August\or September\or October\or November\or December\fi
 \space\number\day, \number\year}

%
\def\chaphead{}		
\newcount\eqnumber
\eqnumber=0		

   \def\neweq {\global\advance\eqnumber by 1%
               \eqno(\hbox{\chaphead\the\eqnumber})}
   \def\neweqa{\global\advance\eqnumber by 1%
               \eqno(\hbox{\chaphead\the\eqnumber a})}
   \def\sameeqb{\eqno(\hbox{\chaphead\the\eqnumber b})}
   \def\sameeq#1{(\hbox{\chaphead\the\eqnumber #1})}
   \def\eqnam#1{\global \advance\eqnumber by 1
                \xdef#1{(\chaphead\the\eqnumber} \relax
 	        \immediate\write1{\def\string#1{#1}} 
 	        \global \advance\eqnumber by -1 
 	      }
   \def\refeq#1{\advance\eqnumber by -#1 \advance\eqnumber by 1
	        (\chaphead\the\eqnumber
     		\advance\eqnumber by #1 \advance\eqnumber by -1 }  %
\def\newe{\global\advance\eqnumber by 1(\hbox{\chaphead\the\eqnumber})}
\def\firste{\global\advance\eqnumber by 1(\hbox{\chaphead\the\eqnumber a})}
\def\laste#1{(\hbox{\chaphead\the\eqnumber #1})}
\def\refe#1{\refeq#1}
%

%
\input mathmacs.tex

\input refmacs.tex

%% file: mathmacs.tex
%
%
%
\def\getfig#1{\ifpsfiles\psfig{figure=#1,width=\hsize}\fi}
%
%
%
%
 
\newcount\fignum
\fignum = 0
\def\figno{\global \advance\fignum by 1 \the\fignum}
\def\thefigno{\the\fignum}
\def\figname#1{\global\advance\fignum by 1%
\xdef#1{\the\fignum}%
\the\fignum%
\relax\immediate\write1{\def\string#1{#1}}}
%
%
\font\fivebmi=cmmib6
\font\sixbmi=cmmib6	\skewchar\sixbmi='177
\font\ninebmi=cmmib10 at 9pt 	\skewchar\ninebmi='177
\newfam\bmifam
\textfont\bmifam=\ninebmi
\scriptfont\bmifam=\sixbmi
\scriptscriptfont\bmifam=\fivebmi
\def\bmi{\fam\bmifam\ninebmi}
\def\b#1{{\bmi#1}}


\mathchardef\alpha="710B
\mathchardef\beta="710C
\mathchardef\gamma="710D
\mathchardef\delta="710E
\mathchardef\epsilon="710F
\mathchardef\zeta="7110
\mathchardef\eta="7111
\mathchardef\theta="7112
\mathchardef\iota="7113
\mathchardef\kappa="7114
\mathchardef\lambda="7115
\mathchardef\mu="7116
\mathchardef\nu="7117
\mathchardef\xi="7118
\mathchardef\pi="7119
\mathchardef\rho="711A
\mathchardef\sigma="711B
\mathchardef\tau="711C
\mathchardef\upsilon="711D
\mathchardef\phi="711E
\mathchardef\chi="711F
\mathchardef\psi="7120
\mathchardef\omega="7121
\mathchardef\varepsilon="7122
\mathchardef\vartheta="7123
\mathchardef\varpi="7124
\mathchardef\varrho="7125
\mathchardef\varsigma="7126
\mathchardef\varphi="7127
 
\def\i{\relax\ifmmode{\rm i}\else\char16\fi}
\def\e{{\rm e}}
%
%

%
\def\spose#1{\hbox to 0pt{#1\hss}}
\def\lta{\mathrel{\spose{\lower 3pt\hbox{$\mathchar"218$}}
     \raise 2.0pt\hbox{$\mathchar"13C$}}}
\def\gta{\mathrel{\spose{\lower 3pt\hbox{$\mathchar"218$}}
     \raise 2.0pt\hbox{$\mathchar"13E$}}}

%
\def\s#1{\widetilde{#1}}
\def\=#1{\overline{#1}}

%

%

%

\def\frac#1#2{{#1 \over #2}}
\def\sfrac#1#2{\leavevmode\kern.1em
  \raise.5ex\hbox{\the\scriptfont0 #1}\kern-.1em
  /\kern-.15em\lower.25ex\hbox{\the\scriptfont0 #2}}

%
\def\d{{\rm d}}

\def\dddot#1{\ddot#1\kern-1.4pt\dot{\phantom{#1}}\kern-3pt}

%


\def\etal{{\it et al.\ }}

\def\={\overline}
\def\s{\ifmmode \widetilde \else \j\fi} 

\def\deg{^\circ}             
\def\micron{\,\mu\hbox{m}}

\def\kms{{\rm\,km\,s^{-1}}}

\def\pc{{\rm\,pc}}
\def\kpc{{\rm\,kpc}}

\def\msun{{\rm\,M_\odot}}

\def\Gyr{{\rm\,Gyr}}

%% file: refmacs.tex
%
\def\etal{{\it et~al.\ }}
%
%
\newcount\refnum

\def\ref{\par\hangindent=1.0cm\hangafter=1}
\def\refn{\par\hangindent=1.0cm\hangafter=1 
	\global \advance\refnum by 1 [\the\refnum]\ }
\def\refnstar{\par\hangindent=1.0cm\hangafter=1 
	\global \advance\refnum by 1 [\the\refnum$^\ast$]\ }
%
\def\aa#1#2{A\&A\ {\bf #1}, #2}

\def\aj#1#2{AJ {\bf #1}, #2}

\def\apj#1#2{ApJ {\bf #1}, #2}

\def\mn#1#2{MNRAS {\bf #1}, #2}

\def\iau127#1{in de Zeeuw P.T. ed, Structure and Dynamics of 
     Elliptical Galaxies, IAU Symp.~No.~127. Reidel, Dordrecht, p.~#1}

%
%

\def \endrefs {\par \endgroup}
%